\documentclass[openacc]{article} 

\usepackage{amsmath,amsfonts,amssymb}
\usepackage{graphicx}
\usepackage{bm} 
\usepackage{tabularx}
\usepackage{comment}
\usepackage{graphicx}
\usepackage{bm} 
\usepackage{subcaption}
\usepackage[a4paper,margin=2.7cm]{geometry}
\usepackage[utf8]{inputenc}
\usepackage[T1]{fontenc}
\renewcommand{\Im}[1]{ \mathrm{Im}{#1} }

\newcommand{\sign}{\text{sgn}}

\def\vph{{\bm \varphi}}

\def\hK{\widehat{\bm{ \mathcal K}}}
\def\K{\bm { \mathcal K}}

\def\B{\bm{ \mathcal B}}

\def\E{\mathcal E}

\title{Semiclassical Analysis of Tunneling in Graphene under Nonuniform Electrostatic and Magnetic Fields}

\author{Maria V. Perel}
\date{}

\begin{document}
	
	\maketitle
	
\begin{center}
	Department of Higher Mathematics and Mathematical Physics,\\
	St Petersburg State University,\\
	7/9 Universitetskaya nab., St Petersburg 199034, Russian Federation\\[2mm]
	E-mail: \texttt{m.perel@spbu.ru}
\end{center}

\bigskip
\noindent\textbf{Keywords:
Dirac equation; semiclassical analysis; nonadiabatic transitions;
quantum tunnelling; graphene; transfer matrix; Fabry--Pérot resonances.}
	
	\begin{abstract}
We develop a semiclassical theory of tunnelling of Dirac fermions through an $n$–$p$–$n$ junction in monolayer graphene subjected to a perpendicular magnetic field.  Electrostatic and magnetic fields are assumed to be smooth functions of a single spatial coordinate, supported on a finite interval, vanishing outside it, and thus ensuring asymptotically free states. In contrast to earlier studies restricted to constant magnetic fields and exactly solvable electrostatic potential profiles, we consider a general electrostatic potential forming an $n$–$p$–$n$ junction and an arbitrary magnetic field, and formulate the corresponding scattering problem.
Within the semiclassical approximation, and under an additional assumption on the incidence angle, the problem reduces to a connection problem for a pair of coalescing turning points, treated using results from our earlier work. We obtain explicit expressions for the reflection and transmission coefficients, including their phases, as functions of energy and incidence angle. Furthermore, we derive semiclassical conditions for Fabry--P\'erot resonances and “magic” angles, and analyse the resulting interference pattern. Numerical results demonstrate the angular dependence of transmission induced by the magnetic field.
 	\end{abstract}

	\section{Introduction}
	
		The ability to control the transport of charge carriers (electrons and holes) in graphene using external electromagnetic fields is crucial for the development of electronic devices. Owing to   weak particle–particle interactions, the behaviour of individual carriers can be  described within a single-particle framework \cite{novoselov2004electric, novoselov2005two, neto2009electronic, katsnelson2007graphene, beenakker2008colloquium, ferrari2015science, katsnelson2020physics}. At low energies, charge carriers obey a 
		 two-dimensional massless Dirac equation with a two-component spinor structure \cite{neto2009electronic, beenakker2008colloquium}, neglecting intervalley scattering. External electrostatic and magnetostatic fields enter this description through scalar and vector potentials.
	
	The study of charge-carrier dynamics in electrostatic potentials led to the discovery of Klein tunnelling \cite{katsnelson2006chiral, katsnelson2020physics}, named by analogy with the relativistic effect \cite{klein1929reflexion}: Charge carriers incident normally on a one-dimensional potential barrier are transmitted with unit probability, regardless of the barrier height. This implies that purely electrostatic fields cannot confine carriers along their direction of variation.   For oblique incidence, however, partial reflection occurs. This regime has been analysed for several exactly solvable potentials, including linear profiles forming $n$–$p$ junctions \cite{cheianov2006selective}, as well as trapezoidal and hyperbolic-secant barriers \cite{sonin2009effect, hartmann2010smooth}. Arbitrary smooth potentials generating 
	$n$–$p$–$n$ junctions were treated asymptotically in \cite{Tudorovskiy_2012, reijnders2013semiclassical} for massless fermions and in \cite{zalipaev2015resonant} for massive fermions. 
	These studies also demonstrated the emergence of Fabry–Pérot resonances due to multiple reflections at the interfaces.

In contrast, inhomogeneous magnetic barriers without electrostatic potentials were considered in	 \cite{de2007magnetic, dell2009multiple}, and methods for generating  inhomogeneous magnetic fields were discussed in \cite{downing2016massless}.

Some qualitative features of carrier transport in the presence of both electrostatic and magnetic fields were discussed in  \cite{cheianov2006selective, shytov2007transport, shytov2009atomic, shytov2008klein}. 
	 Suppression of transport in 
	$n$–$p$–$n$ junctions by magnetic fields was  predicted in \cite{cheianov2006selective}: the curvature of carrier trajectories effectively modifies the incidence angle at the interfaces, thereby reducing transmission. In \cite{shytov2007transport, shytov2009atomic}, transmission coefficients for 
	$n$–$p$ junctions in constant electric and magnetic fields were obtained using Lorentz transformations, revealing two distinct regimes: weak magnetic fields, which perturb tunnelling, and strong fields, which alter the Landau-level structure. Interference effects for a parabolic barrier in a constant magnetic field were analysed in \cite{shytov2008klein} under the assumption that both the electrostatic potential and the vector potential are symmetric.

	Quantitative analyses of carrier transport in both electric and magnetic fields have mostly focused on constant magnetic fields and potentials that allow exact solutions, which are often expressed in analytically involved forms and require further numerical evaluation. Electron transmission through rectangular and double barriers was studied in \cite{masir2010fabry}, whereas triangular barriers were investigated numerically in \cite{mekkaoui2022tunneling}. Similar problems have also been considered in other Dirac materials \cite{kong2021oblique}. Transport in weak magnetic fields and stepped potentials was analysed in \cite{carmier2011semiclassical}.

Despite this progress, the scattering problem for Dirac fermions in the presence of simultaneously varying electrostatic and magnetic fields of arbitrary smooth profiles—forming 
$n$–$p$ and $p$–$n$ junctions and vanishing outside a finite interval—has not yet been  systematically addressed.
	 To  our knowledge, explicit semiclassical expressions for both the amplitudes and phases of the scattering coefficients  for such combined fields are  not available.
	
	In this paper, we present a  study of stationary transport of two-dimensional Dirac fermions in graphene through smooth electrostatic and magnetic barriers coexisting within the same spatial region.  No symmetry assumptions are imposed.
	We adopt a gauge in which both the scalar and vector potentials depend solely on a single spatial coordinate. 
	The electrostatic potential forms a $n$–$p$–$n$ junction with monotonically increasing and decreasing regions separated by a plateau, while the magnetic field is perpendicular, weak according to the criterion of \cite{shytov2007transport, shytov2009atomic}, and features a single maximum. Outside a finite interval, both fields vanish.
 By solving the corresponding scattering problem, we determine both the amplitudes and phases of the reflection and transmission coefficients, providing explicit expressions suitable for qualitative and quantitative analysis of carrier transport in these  barrier configurations.

	Our approach is based on a semiclassical construction of transfer matrices associated with individual $n$–$p$ and $p$–$n$ interfaces. The scattering problem for the Dirac equation is closely related to the Landau–Zener problem, as noted in \cite{shytov2008klein, sonin2009effect}. In this analogy, the spatial coordinate replaces time, while the spatially varying potential produces an avoided crossing of the real-valued, position-dependent longitudinal momentum, with a gap separating the electron and hole branches.  However, significant differences nevertheless emerge. The Dirac equation can be expressed as a Schrödinger-type equation with a non-selfadjoint matrix Hamiltonian that admits a specific factorization.
	 Similar formulations arise in wave-propagation problems, including electromagnetic waves in the Earth–ionosphere waveguide \cite{PerelRadiophys} and elastic waves in inhomogeneous waveguides \cite{perel2005asymptotic}. A general framework for such scattering problems was developed in \cite{FialkovskyPerel} and we apply here the general results  to the case of Dirac fermions.

	The phase of the reflection coefficient plays a crucial role in determining interference effects in the 
	$n$–$p$–$n$ structure.  In particular, it enables us to derive explicit semiclassical conditions for Fabry–Pérot resonances as well as “magic” angles corresponding to enhanced transmission.  In the absence of a magnetic field, our results reduce to previously obtained semiclassical expressions \cite{Tudorovskiy_2012, reijnders2013semiclassical}.
	
	We further predict characteristic restructuring of the interference pattern at energies and angles corresponding to perfect transmission through individual interfaces. Finally, numerical examples are presented to illustrate the significant influence of the magnetic field on carrier transmission through the junction.

	\section{Model and Basic Equations}\label{sec2}
	
\begin{figure}
		\centering
		
		\begin{subfigure}{0.75\textwidth}
			\centering
			\includegraphics[width=\textwidth]{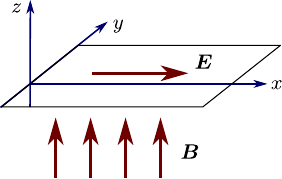}
			\caption{Geometry and field configuration.}
		\end{subfigure}
		
		\vspace{1em}
		
		\begin{subfigure}{0.48\textwidth}
			\centering
			\includegraphics[width=\textwidth,height=5cm]{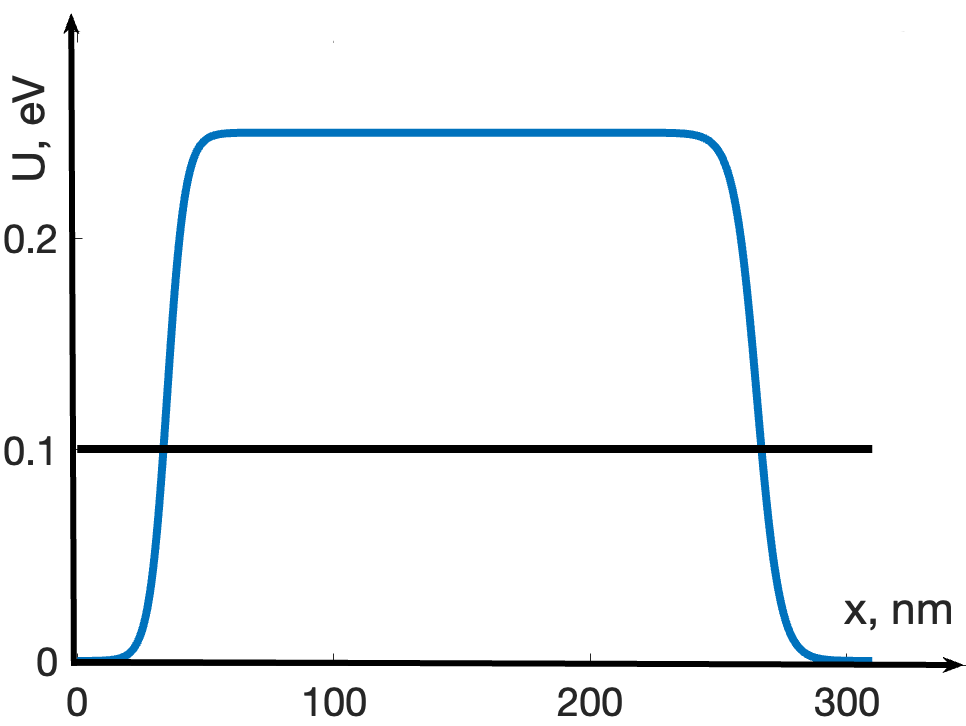}
			\caption{Electrostatic barrier \( U(x) \).}
		\end{subfigure}
		\hfill
		\begin{subfigure}{0.48\textwidth}
			\centering
			\includegraphics[width=\textwidth,height=5cm]{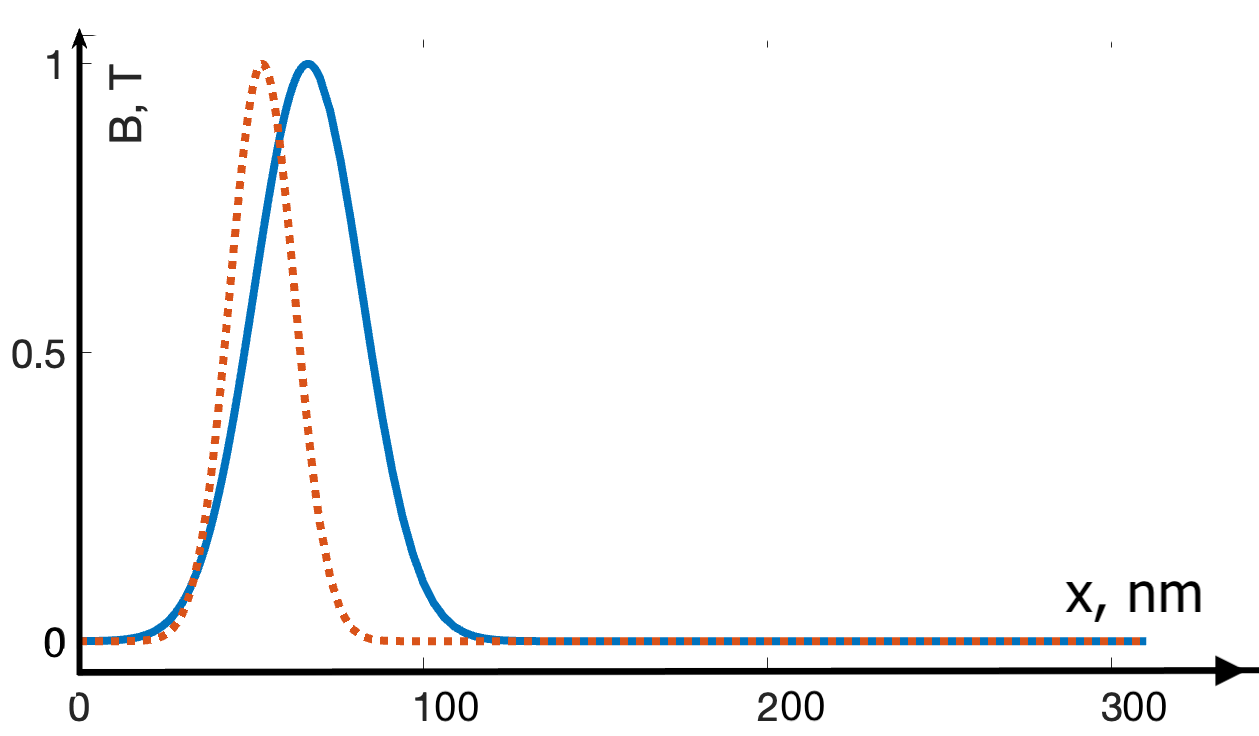}
			\caption{Magnetic-field profiles \( B(x) \).}
		\end{subfigure}
		
		\caption{Schematic of Klein tunnelling in graphene under external fields. The horizontal line indicates the carrier energy \( \mathcal{E} \).}
		\label{Fig1}
	\end{figure}

	We consider a graphene monolayer subjected to external static electric and magnetic fields,
	$\mathbf{E}$ and $\mathbf{B}$, respectively. The field configuration is shown in
	figure~\ref{Fig1}. A graphene sheet lies in the $(x,y)$ plane. The electric field 
	is directed along the 
	$x$ axis (or opposite to it), while the magnetic field is perpendicular to the graphene plane. The electrostatic potential varies within the interval $x_*<x<x_{**}$.
	Both fields depend smoothly on $x$ and are uniform
	in the $y$ direction.
	
	The fields are described by a scalar potential $U(x)$ and a vector potential
	$\mathbf{A}(x)$, defined up to a gauge transformation. Choosing the Landau gauge
	\cite{landau1930diamagnetismus},
	\begin{equation}
		\mathbf{A}=(0,A_y(x),0),
	\end{equation}
	the vector potential is fixed up to an additive constant,
	\begin{equation}\label{eq:Ay0}
		A_y(x)=A_y(0)+\int_0^x B(\tilde{x})\,d\tilde{x}.
	\end{equation}
	Any other gauge choice would introduce an explicit 
	$y$-dependence of 
$A_y$.
	The corresponding fields read
	\begin{equation}
		\mathbf{B}=\nabla\times\mathbf{A}, \qquad
		\mathbf{E}=-\nabla U(x),
	\end{equation}
	where 	$\mathbf{B}=(0,0,B(x)).$
	
	The dynamics of a single charge carrier is governed by the $(2+1)$-dimensional
	Dirac equation \cite{neto2009electronic},
	\begin{equation}
		i\hbar\frac{\partial}{\partial t}\boldsymbol{\Upsilon}
		=\hat{H}(\hat{\mathbf{p}},\mathbf{r})\boldsymbol{\Upsilon},
		\qquad \mathbf{r}=(x,y),
	\end{equation}
	with the Hamiltonian
	\begin{equation}\label{eq:Dirac-K-point}
		\hat{H}
		= v_F\,\boldsymbol{\sigma}\!\cdot\!
		\left(\hat{\mathbf{p}}+e\mathbf{A}(\mathbf{r})\right)
		+ U(\mathbf{r}),
	\end{equation}
	where $\hat{\mathbf{p}}=-i\hbar\nabla$, $v_F$ is the Fermi velocity, and $e$ is the
	electron charge. The Pauli matrices are
	\begin{equation}
		\sigma_x=
		\begin{pmatrix}
			0 & 1\\
			1 & 0
		\end{pmatrix},
		\qquad
		\sigma_y=
		\begin{pmatrix}
			0 & -i\\
			i & 0
		\end{pmatrix},
	\end{equation}
	and $\boldsymbol{\sigma}=(\sigma_x,\sigma_y)$. The Hamiltonian
	\eqref{eq:Dirac-K-point} describes carriers near the $K$ point of the graphene
	band structure.
	We restrict ourselves to weak magnetic fields, defined by the condition
	\begin{equation}\label{eq:weak-field}
		\bigl|v_F^{-1}U'(x)\bigr| > e|B(x)|.
	\end{equation}
	For a locally homogeneous system with the same values of $U'(x)$ and $B(x)$,
	condition~\eqref{eq:weak-field} implies that a Lorentz frame exists in which the
	magnetic field vanishes \cite{shytov2007transport}. In this case,  the electric field dominates over the magnetic field in the semiclassical dynamics. This condition must hold at
	the  turning points defined by $\mathcal{E}=U(x)$. The carrier energy
	$\mathcal{E}$ lies in the interval $(0,\max U)$ and is assumed not to be close to
	$\max U$.
	
	We seek stationary states of energy $\mathcal{E}$ in the form
	\begin{equation}\label{Upsilon-Psi}
		\boldsymbol{\Upsilon}(\mathbf{r},t)
		= e^{i(p_y y-\mathcal{E}t)/\hbar}\,\boldsymbol{\Psi}(x),
	\end{equation}
	where $p_y$ is the conserved transverse momentum. Substitution into the Dirac
	equation yields a one-dimensional system,
	\begin{equation}\label{eq:one-dimensionalDirac}
		-i\hbar\sigma_x\frac{d\boldsymbol{\Psi}}{dx}
		= \hK(x)\boldsymbol{\Psi},
		\qquad
		\hK(x)=
		\begin{pmatrix}
			\mathcal{E}-U(x) & i {\cal P}\\
			-i {\cal P} & \mathcal{E}-U(x)
		\end{pmatrix},
	\end{equation}
	with
	\begin{equation}\label{eq:P}
		{\cal P}=p_y+eA_y(x).
	\end{equation}
	
	We introduce dimensionless variables according to
	\begin{equation}\label{eq:dim}
		x=\frac{\check{x}}{\check{l}},\quad
		\mathcal{E}=\frac{\check{\mathcal{E}}}{\check{v}_F\check{p}_0},\quad
		U=\frac{\check{U}}{\check{v}_F\check{p}_0},\quad
		A_y=\frac{\check{e}\check{A}_y}{\check{p}_0},\quad
		p_y=\frac{\check{p}_y}{\check{p}_0},\quad
		\hbar=\frac{\check{\hbar}}{\check{p}_0\check{l}}.
	\end{equation}
	Quantities with a check denote dimensional variables.
	 We adopt the values
	$\check{v}_F=10^6\,{\rm m s}^{-1}$, $\check{\hbar}=6.6\times10^{-16}\,$eV$\cdot$s,
	and $\check{e}=1\,$eV/V.
	
	As a representative example, we consider an electrostatic barrier with
	$\max\check{U}=250\,$meV, $\check{l}_{n\text{--}p}=70\,$nm,
	$\check{l}_{p\text{--}n}=90\,$nm, and a constant-potential region of length
	$\check{l}_{\mathrm{const}}=150\,$nm, as in \cite{reijnders2013semiclassical}.
	Choosing $\check{\mathcal{E}}_0-\check{U}_0=0.1\,$eV and $\check{l}=70\,$nm, we define
	\[
	\check{p}_0=(\check{\mathcal{E}}_0-\check{U}_0)\check{v}_F^{-1},
	\]
	which gives
	\begin{equation}\label{eq:hbar-app}
		\hbar=\frac{\check{\hbar}\check{v}_F}
		{(\check{\mathcal{E}}_0-\check{U}_0)\check{l}}
		\approx 0.1.
	\end{equation}
	
	In what follows, all dimensionless variables and their derivatives are assumed to
	be of order unity. We set 
	$e=v_F=1$ and omit them hereafter.  The small parameter $\hbar\ll1$
	 serves as the semiclassical expansion parameter.
	
		\section{Semiclassical Approach}\label{sec3}
	
	We solve equation \eqref{eq:one-dimensionalDirac} using the semiclassical (WKB)
	approximation \cite{berry72,landau2013quantum}. In this framework, the solution is
	expressed in terms of the local classical momentum. As long as the semiclassical
	approximation remains valid, a charge carrier follows a single classical
	trajectory and no scattering occurs. Scattering becomes possible only in regions
	where the semiclassical approximation breaks down.
	
	Mathematically, equation~\eqref{eq:one-dimensionalDirac} is a system
	of ordinary differential equations with a small parameter $\hbar$ multiplying the
	derivative. Semiclassical methods for such systems are well established
	\cite{fedoryuk2012asymptotic,maslov2001semi}. In the absence of a magnetic field,
	this equation was analyzed asymptotically in
	 \cite{Tudorovskiy_2012,reijnders2013semiclassical,zalipaev2015resonant} by
	reducing it to a second-order scalar equation. In contrast, we treat the
	first-order system directly, which allows for a transparent inclusion of the magnetic
	field.
	
	We employ the semiclassical (WKB) ansatz
	\begin{equation}\label{f:one-dimensionalsolution}
		\boldsymbol{\Psi}(x) =
		\exp\Big[\frac{i}{\hbar} \int^x \beta(s)\, ds \Big]
		\sum_{m=0}^{\infty} \hbar^m \boldsymbol{\Phi}^{(m)}(x),
	\end{equation}
	where $\beta(x)$ and $\boldsymbol{\Phi}^{(m)}(x)$ are sufficiently smooth functions of $x$.
	Substitution of \eqref{f:one-dimensionalsolution} into
	\eqref{eq:one-dimensionalDirac} yields a hierarchy of algebraic equations.
	
	The zeroth-order equation gives
	\begin{equation}\label{eq:Phi0}
		\boldsymbol{\Phi}^{(0)}_j(x) = C^{(0)}_j(x)\, \boldsymbol{\varphi}_j(x),
	\end{equation}
	where $\boldsymbol{\varphi}_j(x)$ satisfies
	\begin{equation}\label{eq:eig-un}
		\hK(x)\, \boldsymbol{\varphi}_j(x) = \beta_j(x)\, \sigma_x \boldsymbol{\varphi}_j(x).
	\end{equation}

	The eigenvalues $\beta_j(x)$ are given by appropriate branches of
	\begin{equation}\label{eq:beta}
		\beta(x) = \sqrt{(\E-U(x))^2 - {\cal P}^2(x)},
	\end{equation}
	with the choice of branches specified below. They represent the longitudinal momentum, i.e. the classical momentum along the $x$-direction.
	
	Regions where $\beta(x)$ is real (imaginary) correspond to classically allowed (forbidden) motion. Turning (degeneracy) points $\varkappa$ for linear systems of ordinary differential equations are defined as the coordinates $x$ at which two eigenvalues $\beta$ coincide, see for example \cite{fedoryuk2012asymptotic}. For system \eqref{eq:one-dimensionalDirac}, the turning points satisfy the condition $\beta(\varkappa)=0$.
	
	If ${\cal P}(x)=0$, there is a single doubly degenerate turning point $\varkappa_{\cal E}$ determined by
	\begin{equation}\label{eq:e-turn-p}
		{\cal E} = U(\varkappa_{\cal E}),
	\end{equation}
	which corresponds to Klein tunnelling \cite{katsnelson2006chiral}.
	
	For ${\cal P}(\varkappa_{\cal E}) \neq 0$, this degeneracy is lifted, and the turning point splits into two simple turning points $\varkappa_- < \varkappa_+$ defined by
	\begin{equation}\label{eq:deg-p}
		|\E - U(\varkappa_{\pm})| = |{\cal P}(\varkappa_{\pm})|.
	\end{equation}
	
	In the classically allowed regions, the eigenvalues are labeled as follows:
	\begin{equation}\label{eq:betas}
		\beta_1(x) = -\beta_2(x), \qquad
		\beta_2(x) =
		\begin{cases}
			|\beta(x)|, & x > \varkappa_+,\\
			-|\beta(x)|, & x < \varkappa_-.
		\end{cases}
	\end{equation}
	The corresponding eigenvectors take the form
	\begin{equation}\label{eq:varphi}
		\boldsymbol{\varphi}_j(x) =
		\begin{pmatrix}
			\beta_j(x) - i {\cal P}(x)\\[1ex]
			\E - U(x)
		\end{pmatrix}.
	\end{equation}
	
	The solvability condition for the first-order correction
	$\boldsymbol{\Phi}_j^{(1)}(x)$ leads to
	\begin{equation}\label{eq:conver}
		C^{(0)}_j(x) =
		\frac{\mathcal{A}_j}{\sqrt{|N_j(x)|}}
		\exp\Big[- i\int_a^x \Im S_j(x')\, dx'\Big],
	\end{equation}
	where
	\begin{equation}\label{eq:S-N}
		S_j(x) = \frac{(\boldsymbol{\varphi}_j, \sigma_x \boldsymbol{\varphi}'_j)}{(\boldsymbol{\varphi}_j, \sigma_x \boldsymbol{\varphi}_j)}, \qquad
		N_j(x) = (\boldsymbol{\varphi}_j, \sigma_x \boldsymbol{\varphi}_j) = 2 \beta_j(x) (\E-U(x)).
	\end{equation}
	
	Evaluating the integral in \eqref{eq:conver} yields the leading-order
	semiclassical solution
	\begin{equation}\label{eq:adiab}
		\boldsymbol{\Psi}_j(x) \approx
		\frac{\boldsymbol{\varphi}_j(x)}{\sqrt{|N_j(x)|}}
		\exp\Bigg\{
		\frac{i}{\hbar} \int_a^x \beta_j(x')\, dx' - i \gamma_j(x) + i\gamma_j(a)\,
		\Bigg\},
	\end{equation}
	where the Berry phase is given by
	\begin{equation}\label{eq:Berry}
		\gamma_j(x) = -\frac{1}{2} \arcsin \frac{{\cal P}(x)}{|\E-U(x)|} \,
		\mathrm{sgn}\beta_j(x).
	\end{equation}
	
	Choosing the lower limit $a=\varkappa_c$ as the nearest turning point $\varkappa_c = \varkappa_\pm$, we obtain
	\begin{multline}\label{eq:adiab-last}
		\boldsymbol{\Psi}_j^{\,c}(x) =
		\frac{\boldsymbol{\varphi}_j(x)}{\sqrt{|2 \beta_j(x) (\E-U(x))|}}\\
		\times \exp\Bigg[
		i\, \mathrm{sgn}\beta_j(x) \Big(
		\frac{1}{\hbar} \int_{\varkappa_c}^{x} |\beta_j(x')|\, dx' +
		\frac{1}{2} \arcsin \frac{{\cal P}(x)}{|\E-U(x)|}
		\Big)\, + i\,	\gamma_{j}(\varkappa_c)\,
		\Bigg]
	\end{multline}
	Here
	\begin{equation}\label{eq:gamma-E}
	c =\pm, \quad	\gamma_{j}(\varkappa_c) = -\mathrm{sgn}\beta_j(x)\,\chi_{\E},\quad \varkappa_c=\varkappa_{\pm}, \quad \chi_{\E} = \frac{\pi}{4} \mathrm{sgn} {\cal P}(\varkappa_{\E}), 
	\end{equation}
	which results from the assumption of  weak $B$ and the relations
	\begin{equation}
		\begin{aligned}
			&\varkappa_c - \varkappa_{\E} \approx 
			\pm\frac{p_y + A_y(\varkappa_{\E})}{U'(\varkappa_{\E})  \mp B(\varkappa_{\E})},\\	
			&{\cal P}(\varkappa_c)  \approx  {\cal P}(\varkappa_{\E}) + B(\varkappa_{\E})\left(  \varkappa_c - \varkappa_{\E}\right) =
			{\cal P}(\varkappa_{\E})
			\frac{U'(\varkappa_{\E})}{U'(\varkappa_{\E})  \mp B(\varkappa_{\E})}.
		\end{aligned} 
	\end{equation}

	The approximation \eqref{eq:adiab-last} breaks down at the turning
	points $x=\varkappa_{\pm}$, where $\beta_j(x)$ vanishes.
	A general solution can be written as
	\begin{equation}
		\boldsymbol{\Psi}(x) \approx
		\begin{cases}
			k_{1}^{-} \boldsymbol{\Psi}_{1}^{-}(x) + k_{2}^{-} \boldsymbol{\Psi}_{2}^{-}(x), & x \ll \varkappa_-,\\
			k_{1}^{+} \boldsymbol{\Psi}_{1}^{+}(x) + k_{2}^{+} \boldsymbol{\Psi}_{2}^{+}(x), & x \gg \varkappa_+,
		\end{cases}
	\end{equation}
	which defines the transfer matrix $\mathcal{T}$:
	\begin{equation}\label{eq:T-defin}
		\begin{pmatrix} k_1^{+} \\ k_2^{+} \end{pmatrix} =
		\mathcal{T} \begin{pmatrix} k_1^{-} \\ k_2^{-} \end{pmatrix}.
	\end{equation}
	
		\section{Scattering Problem}
	
\begin{figure}[!htbp]
		\includegraphics[width=.9\textwidth]{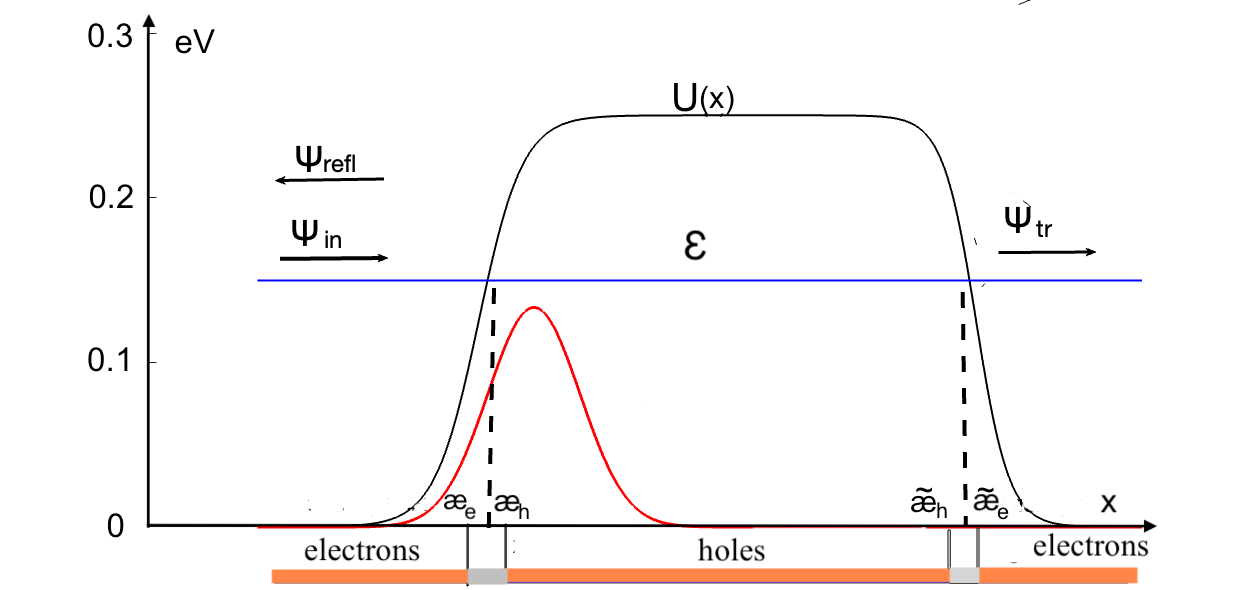}
		\caption{Charge carrier scattering at a potential barrier $U(x)$ in a spatially varying magnetic field $B(x)$.   Horizontal line: carrier energy 
			$\mathcal{E}$. Incoming, reflected, and transmitted waves are $\boldsymbol{\Psi}_{in}$, $\boldsymbol{\Psi}_{refl},$ and $\boldsymbol{\Psi}_{tr}.$ 
			Dashed vertical lines indicate $x=\varkappa_{\mathcal E}$, defined by ${\mathcal E} = U(\varkappa_{\mathcal E})$. Small ticks indicate the turning (degeneracy) points $\varkappa_{e}$, $\varkappa_{h}$, $\tilde{\varkappa}_{h}$, and $\tilde{\varkappa}_{e}$, which delimit the interfaces between classically allowed and forbidden regions. The central allowed region is hole-like, whereas the outer allowed regions are electron-like. }
		\label{fig:scattering}
	\end{figure}

	We consider scattering of a charge carrier by an electrostatic potential barrier in the presence of an external magnetic field (see figure~\ref{fig:scattering}).
	
	The charge carrier is incident on the barrier at an angle $\theta$ with energy $\mathcal{E}$. We assume that the electromagnetic field vanishes for
	   $x \le x_*$, and choose the gauge $A_y(x_*)=0$, which affects the wave function only through an irrelevant 
	   $y$-dependent phase.
	From \eqref{eq:beta}, at $x = x_*$ one has
	\begin{equation}
		\beta^2(x_*) + p_y^2 = \mathcal{E}^2,
	\end{equation}
	so that
	\begin{equation}\label{eq:py}
		\beta(x_*) = \mathcal{E} \cos\theta, \qquad
		p_y = \mathcal{E} \sin\theta.
	\end{equation}
	For $x > x_*$ the function $\beta(x)$ is determined by equation \eqref{eq:beta}.
	We also assume that fields vanish for $x \ge x_{**}$.
	
	The wave function is written as
	\begin{equation}
		\boldsymbol{\Psi}(x) =
		\begin{cases}
			\boldsymbol{\Psi}_{\mathrm{in}}(x) + r_{\text{n-p-n}} \,\boldsymbol{\Psi}_{\mathrm{refl}}(x), & x \le x_*, \\
			t_{\text{n-p-n}} \,\boldsymbol{\Psi}_{\mathrm{tr}}(x), & x \ge x_{**}.
		\end{cases}
	\end{equation}
	Here $\boldsymbol{\Psi}_{\mathrm{in}}$ and $\boldsymbol{\Psi}_{\mathrm{refl}}$ describe incoming and reflected waves to the left of the barrier, while $\boldsymbol{\Psi}_{\mathrm{tr}}$ corresponds to the transmitted wave.
	We denote the  transmitted and reflected coefficients by $t_{\text{n-p-n}}$ and $r_{\text{n-p-n}}$, respectively. 
	
	To construct the transfer matrix, we distinguish classically allowed and forbidden regions.
	For a given energy $\mathcal{E}$, equation \eqref{eq:e-turn-p} has two solutions,
	$\varkappa_{\mathcal{E}}$ and $\tilde{\varkappa}_{\mathcal{E}}$,
	corresponding to increasing and decreasing parts of the potential, respectively.
	Forbidden regions arise in the intervals
	$\varkappa_- < x < \varkappa_+$ and $\tilde{\varkappa}_- < x < \tilde{\varkappa}_+$. 	The semiclassical solutions $\boldsymbol{\Psi}_j^c$ and $\tilde{\boldsymbol{\Psi}}_j^c$ are defined in the vicinity of the corresponding forbidden regions. 
	\begin{figure}[!htbp]
		\includegraphics[width=.35\textwidth]{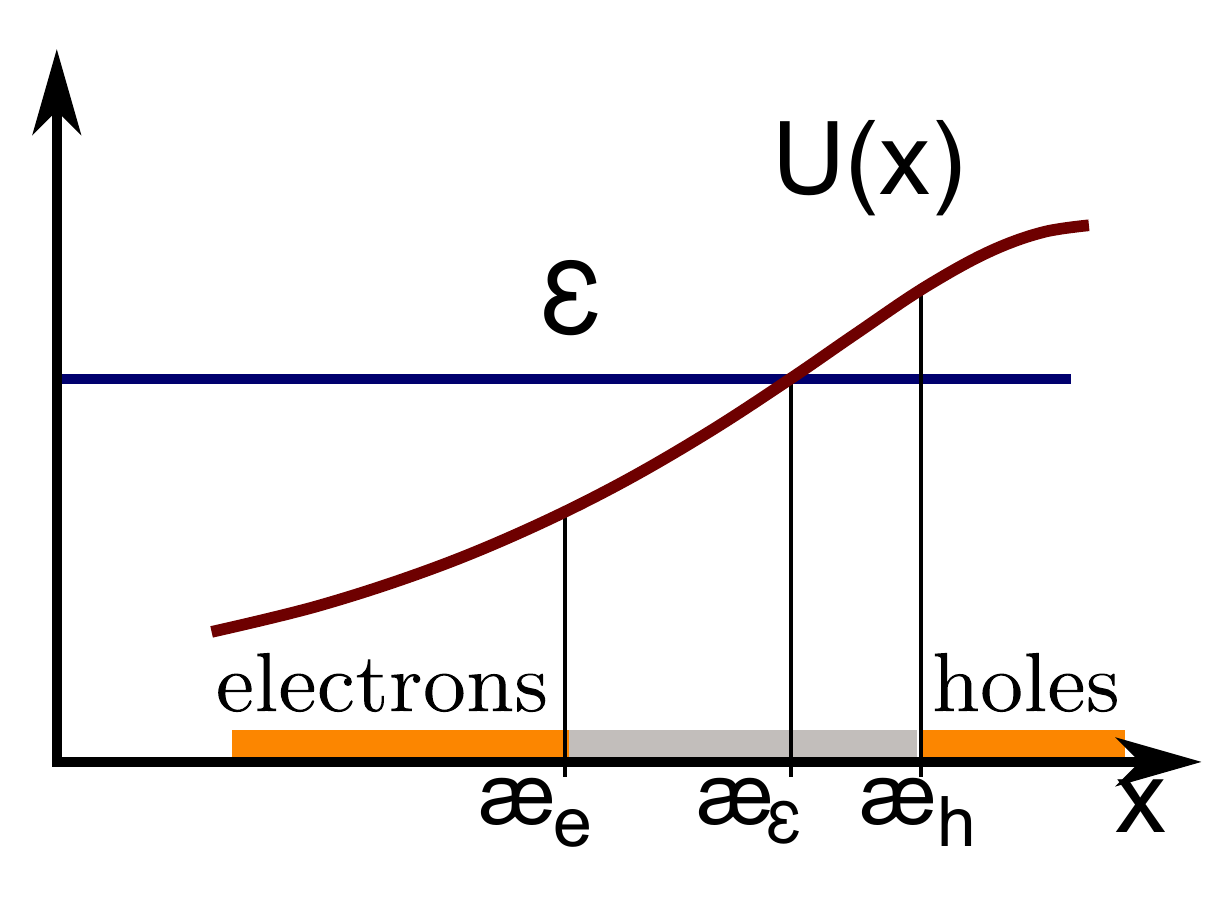}
		\caption{Turning points $\varkappa_e$ and $\varkappa_h$ for electrons and holes with energy ${\cal E}$ in a monotonically increasing electrostatic potential $U(x)$. The classically allowed regions are $x < \varkappa_e$ and $x > \varkappa_h$, whereas the classically forbidden region is $\varkappa_e < x < \varkappa_h$.
			}
		\label{turning points}
	\end{figure}
	The carrier is an electron when $\mathcal{E} > U(x)$ and a hole when $\mathcal{E} < U(x)$, therefore $\varkappa_-\equiv \varkappa_e$, $\varkappa_+\equiv \varkappa_h$, $\tilde\varkappa_-\equiv \tilde\varkappa_h$, $\tilde\varkappa_+\equiv \tilde\varkappa_e$. 
	
	The direction of propagation is determined by the conserved current
	\begin{equation}\label{eq:current-density}
		j_x = (\boldsymbol{\Psi}, \sigma_x \boldsymbol{\Psi}) = \mathrm{const},
	\end{equation}
	which yields
	\begin{equation}\label{eq:current}
		\mathrm{sign}(j_x) = \mathrm{sign}\Bigl( 2 \beta_j(x) (\mathcal{E} - U(x)) \Bigr).
	\end{equation}
	Accordingly, we label the modes by $eR$, $eL$, $hR$, and $hL$.
	
	The correspondence between the semiclassical solutions reads
	\begin{align}
		\boldsymbol{\Psi}_1^{-} &= \boldsymbol{\Psi}_{eR}, & \boldsymbol{\Psi}_2^{-} &= \boldsymbol{\Psi}_{eL}, & \boldsymbol{\Psi}_1^{+} &= \boldsymbol{\Psi}_{hR}, & \boldsymbol{\Psi}_2^{+} &= \boldsymbol{\Psi}_{hL},\label{eq:solutions-corresp} \\
		\tilde{\boldsymbol{\Psi}}_1^{-} &= \tilde{\boldsymbol{\Psi}}_{hL}, & \tilde{\boldsymbol{\Psi}}_2^{-} &= \tilde{\boldsymbol{\Psi}}_{hR}, & \tilde{\boldsymbol{\Psi}}_1^{+} &= \tilde{\boldsymbol{\Psi}}_{eL}, & \tilde{\boldsymbol{\Psi}}_2^{+} &= \tilde{\boldsymbol{\Psi}}_{eR}.
	\end{align}
	For a monotonically increasing potential, the transfer matrix $\mathcal{T}_{\text{n-p}}$ relates the expansion coefficients of a general solution $\boldsymbol{\Psi}(x)$, which has the asymptotic 
	\begin{equation}
		\boldsymbol{\Psi}(x) \approx
		\begin{cases}
			k_{eR} \boldsymbol{\Psi}_{eR}(x) + k_{eL} \boldsymbol{\Psi}_{eL}(x), & x \ll \varkappa_e,\\
			k_{hR} \boldsymbol{\Psi}_{hR}(x) + k_{hL} \boldsymbol{\Psi}_{hL}(x), & x \gg \varkappa_h
		\end{cases}
	\end{equation}
	across the classically forbidden region between the turning points $\varkappa_e$ and $\varkappa_h$:
	\begin{equation}
		\begin{pmatrix}
			k_{hR} \\ k_{hL}
		\end{pmatrix}
		=
		\mathcal{T}_{\text{n-p}}
		\begin{pmatrix}
			k_{eR} \\ k_{eL}
		\end{pmatrix}.
	\end{equation}
	Setting
	\begin{equation}
		k_{eL} = r_{\text{n-p}}, \quad k_{eR} = 1, \quad k_{hL} = 0, \quad k_{hR} = t_{\text{n-p}},
	\end{equation}
	one finds
	\begin{equation}\label{eq:r-t-matrix}
		r_{\text{n-p}} = -\frac{(\mathcal{T}_{\text{n-p}})_{21}}{(\mathcal{T}_{\text{n-p}})_{22}}, \qquad
		t_{\text{n-p}} = \frac{\det \mathcal{T}_{\text{n-p}}}{(\mathcal{T}_{\text{n-p}})_{22}}.
	\end{equation}
	
	For a potential monotonically decreasing along 
	$x$, the transfer matrix ${\cal T}_{p-n} = \tilde{\cal T}$
	relates electron and hole coefficients via
	\begin{equation}
		\begin{pmatrix}
			\tilde{k}_{eL} \\ \tilde{k}_{eR}
		\end{pmatrix}
		=
		\mathcal{T}_{\text{p-n}}
		\begin{pmatrix}
			\tilde{k}_{hL} \\ \tilde{k}_{hR}
		\end{pmatrix},
	\end{equation}
	and, setting $\tilde{k}_{eL}=0$, $\tilde{k}_{eR}=t_{p-n}$, $\tilde{k}_{hL}=r_{p-n}$, $\tilde{k}_{hR}=1$, one finds
	\begin{equation}\label{eq:r-t-defin-pn}
		r_{\text{p-n}} = -\frac{(\mathcal{T}_{\text{p-n}})_{12}}{(\mathcal{T}_{\text{p-n}})_{11}}, \qquad
		t_{\text{p-n}} = \frac{\det \mathcal{T}_{\text{p-n}}}{(\mathcal{T}_{\text{p-n}})_{11}}.
	\end{equation}
	
	Propagation between the two classically forbidden regions is described by the transfer matrix
	\begin{equation}\label{eq:zeta}
		\mathcal{T}_{\text{p-p}} =
		\begin{pmatrix}
			0 & e^{i\zeta} \\
			e^{-i\zeta} & 0
		\end{pmatrix},
		\qquad
		\zeta = \frac{1}{\hbar} \int_{\varkappa_h}^{\tilde{\varkappa}_h} |\beta(x')|\, dx'
		+	\tilde{\chi}_{\E} - \chi_{\E},
	\end{equation}
	where, according to \eqref{eq:gamma-E},
	\begin{equation}
		\tilde{\chi}_{\E} - \chi_{\E} =
		\frac{\pi}{4}
		\left[
		\operatorname{sign}\mathcal{P}(\tilde{\varkappa}_{\E})
		-
		\operatorname{sign}\mathcal{P}({\varkappa}_{\E})
		\right].
	\end{equation}

	The total transfer matrix through the barrier is
	\begin{equation}\label{eq:product-transfer}
		\mathcal{T}_{\text{n-p-n}} = \mathcal{T}_{\text{p-n}} \, \mathcal{T}_{\text{p-p}} \, \mathcal{T}_{\text{n-p}}.
	\end{equation}
	Imposing $\tilde{k}_{eL} = 0$, $\tilde{k}_{eR} = t_{\text{n-p-n}}$, $k_{eR} = 1$, $k_{eL} = r_{\text{n-p-n}}$, we obtain
	\begin{equation}\label{eq:trans-npn}
		r_{\text{n-p-n}} = -\frac{(\mathcal{T}_{\text{n-p-n}})_{11}}{(\mathcal{T}_{\text{n-p-n}})_{12}}, \qquad
		t_{\text{n-p-n}} = -\frac{\det \mathcal{T}_{\text{n-p-n}}}{(\mathcal{T}_{\text{n-p-n}})_{12}}.
	\end{equation}

	\section{Transfer Matrix Calculation \label{sec:transfer}}
	
	In the previous section, we demonstrated that once the transfer matrix
	$\cal T$ is known, the scattering problem for the potential barriers shown in
	Fig.~\ref{fig:scattering} can be solved.
	It has been noted in several works
	(see \cite{shytov2007transport, sonin2009effect})
	that the equations describing charge-carrier scattering by a monotonic
	electrostatic barrier in graphene are closely related to those of the
	Landau--Zener problem.
	A broader class of systems encompassing the present case was analyzed in
	\cite{FialkovskyPerel}, where the distinction from the conventional
	Landau--Zener setting was discussed.
	Here we specialize the transfer matrix obtained in
	\cite{FialkovskyPerel} to our problem. The matrix connects the simplified semiclassical solutions introduced above and differs from the original solutions by a phase factor determined explicitly below.

	\subsection{Reduction of the Dirac Equation to a Schrödinger-Type Equation
		\label{sec:reduction}}
	
	We rewrite the quantity ${\mathcal P}(x)$ appearing in equation
(\ref{eq:one-dimensionalDirac}) as
	\begin{equation}\label{eq:A-decomp-0}
		{\cal P}(x)
		= p_y + A_y(\varkappa_{\E}) + \big[A_y(x) - A_y(\varkappa_{\E})\big],
	\end{equation}
	and assume
	\begin{equation}\label{eq:gamma-def-0}
		p_y + A_y(\varkappa_{\E})
		= \delta \mathcal{P}_{\E},
		\qquad \delta \ll 1,
		\qquad \mathcal{P}_{\E} \sim 1.
	\end{equation}
	This inequality
 thus restricts the allowed charge-carrier incidence angles to values close to the angle of full transmission.
	For a single $p$-$n$ junction, $\varkappa_{\E}$ should be replaced by
	$\tilde{\varkappa}_{\E}$.

	With this decomposition, the matrix $\hK$ in
 (\ref{eq:one-dimensionalDirac}) can be written as
	\begin{equation}\label{for_matrix-0}
		\begin{aligned}
			\hK(x) &= \K(x) + \delta \B, \\
			\K(x) &=
			\begin{pmatrix}
				U(\varkappa_{\E}) - U(x) &
				i\,[A_y(x) - A_y(\varkappa_{\E})] \\
				-i\,[A_y(x) - A_y(\varkappa_{\E})] &
				U(\varkappa_{\E}) - U(x)
			\end{pmatrix}, \\
			\B &=
			\begin{pmatrix}
				0 & i\,\mathcal{P}_{\E} \\
				-i\,\mathcal{P}_{\E} & 0
			\end{pmatrix}.
		\end{aligned}
	\end{equation}
	
	Equation~(\ref{eq:one-dimensionalDirac}) then reduces to
	\begin{equation}\label{eq:oneDirac-FP-0}
		-i\hbar \sigma_x
		\frac{d\boldsymbol{\Psi}(x)}{dx}
		=
		\big[\K(x) + \delta \B\big]\boldsymbol{\Psi}(x).
	\end{equation}
	This equation contains two small parameters: the dimensionless Planck constant
	$\hbar$ and the perturbation parameter $\delta$.
	Formally, it resembles a Schr\"odinger equation with $x$ playing the role of
	time and $\K(x)+\delta\B$ acting as an effective Hamiltonian.
	However, the operator
	$\sigma_x^{-1}[\K(x)+\delta\B]$ is non-Hermitian, and
	$\boldsymbol{\Psi}(x)$ is a two-component vector.
	For this reason, we refer to equation  \eqref{eq:oneDirac-FP-0} as a
	Schr\"odinger-type equation.
	The corresponding scattering problem was analyzed in
	 \cite{FialkovskyPerel}.
	
	We consider the unperturbed eigenvalue problem
	\begin{equation}\label{eq:unpert-0}
		\K(x)\vph_j^0(x)
		=
		\beta_j^0(x)\,\sigma_x\vph_j^0(x),
		\qquad j=1,2,
	\end{equation}
	which has the following properties:
	\begin{itemize}
		\item The eigenvalues $\beta_1^0(x)$ and $\beta_2^0(x)$ cross at
		$x=\varkappa_{\E}$.
		In the vicinity of this point,
		\begin{equation}\label{eq:eigenvalues-difference-0}
			\beta_2^0(x) - \beta_1^0(x)
			\sim 2Q(x-\varkappa_{\E}), \qquad Q>0.
		\end{equation}
		\item The eigenvectors
		$\vph_1^0(\varkappa_{\E})$ and $\vph_2^0(\varkappa_{\E})$
		are linearly independent.
		\item The eigenvalues and eigenvectors are smooth functions of $x$.
	\end{itemize}
	
	Following \cite{FialkovskyPerel}, we assume
	\begin{equation}\label{eq:del-h-0}
		\delta \sim \sqrt{\hbar},
	\end{equation}
	thereby neglecting study of exponentially small contributions to the scattering matrix while retaining nonadiabatic transitions.
	
	In the vicinity of $x=\varkappa_{\E}$, the scattering problem is formulated in terms
	of simplified with account of \eqref{eq:del-h-0} semiclassical solutions named 
	$\boldsymbol{\Psi}_j^{c,0}(x)$,
	\begin{equation}\label{eq:adiab-simpl-0}
		\begin{aligned}
			\boldsymbol{\Psi}_j^{c,0}(x)
			&=
			\frac{\vph_j^0(x)}{|N_j^0(x)|^{1/2}}
			\exp\!\Biggl[
			i \int_{\varkappa_c}^{x}
			\Bigl(
			\frac{\beta_j(x')}{\hbar}
			- \Im S_j^0(x')
			\Bigr)\,dx'
			\Biggr], \\
			N_j^0(x)
			&=
			\bigl(\vph_j^0(x), \sigma_x \vph_j^0(x)\bigr), \\
			S_j^0(x)
			&=
			\frac{
				\bigl(\vph_j^0(x),
				\sigma_x\,d\vph_j^0/dx\bigr)}
			{N_j^0(x)}.
		\end{aligned}
	\end{equation}
	The lower integration limit is $\varkappa_c=\varkappa_{-}$ or $\varkappa_{+}$, depending on the
	region of $x$.
	Unlike the solutions $\boldsymbol{\Psi}_j^{c}(x)$, the amplitude prefactor and
	$S_j^0(x)$ are evaluated using the simplified eigenfunctions
	$\vph_j^0(x)$, whereas the phase contains the exact eigenvalues
	$\beta_j(x)$.
	
	Any solution $\boldsymbol{\Psi}(x)$ can be decomposed as
	\begin{equation}\label{eq:Psi-det-gensimple-0}
		\begin{aligned}
			\boldsymbol{\Psi}(x)
			&\approx
			k_1^{-,0}\boldsymbol{\Psi}_1^{-,0}
			+ k_2^{-,0}\boldsymbol{\Psi}_2^{-,0},
			\qquad x \ll \varkappa_{\E}, \\
			\boldsymbol{\Psi}(x)
			&\approx
			k_1^{+,0}\boldsymbol{\Psi}_1^{+,0}
			+ k_2^{+,0}\boldsymbol{\Psi}_2^{+,0},
			\qquad x \gg \varkappa_{\E}.
		\end{aligned}
	\end{equation}
	The transfer matrix ${\cal T}^0$ is defined by
	\begin{equation}\label{defin-T-gen-0}
		\begin{pmatrix}
			k_1^{+,0} \\
			k_2^{+,0}
		\end{pmatrix}
		=
		{\cal T}^0
		\begin{pmatrix}
			k_1^{-,0} \\
			k_2^{-,0}
		\end{pmatrix},
	\end{equation}
	and takes the form \cite{FialkovskyPerel}
	\begin{equation}\label{eq:transfer-real-v-0}
		{\cal T}^0 =
		\begin{pmatrix}
			e^{\pi|\nu|} &
			e^{2 i \Theta_1}
			\sqrt{|1 - e^{2\pi|\nu|}|} \\
			e^{-2 i \Theta_1}
			\sqrt{|1 - e^{2\pi|\nu|}|} &
			e^{\pi|\nu|}
		\end{pmatrix},
		\qquad \det {\cal T}^0 = 1.
	\end{equation}
	
	The parameter $\nu$ and the phase $\Theta_1$ are given by
	\begin{equation}\label{eq:nu-0}
		\nu = -i \frac{p^2}{2Q}, \qquad
		p^2 =
		\frac{{\cal B}_{12}^{(0)}{\cal B}_{21}^{(0)}}
		{|N_1^0 N_2^0|},
	\end{equation}
	where $N_j^0 = N_j^0(\varkappa_{\E})$ and
	\[
	{\cal B}_{jl}^{(0)} =
	\bigl(\vph_j^0(x), \B(x)\vph_l^0(x)\bigr)
	\big|_{x=\varkappa_{\E}}.
	\]
	The coefficient $Q$ characterizes the linear behavior of the eigenvalues near
	$x=\varkappa_{\E}$, see \eqref{eq:eigenvalues-difference-0}.
	
	The phase $\Theta_1$ reads
	\begin{equation}\label{eq:A-general-0}
		\begin{aligned}
			\Theta_1 &=
			\frac{\theta_a + \theta_\Gamma - \theta_\Gamma^{\rm as}}{2}, \\
			\theta_a &=
			\arg\!\left(
			\frac{{\cal B}_{12}^{(0)}}{N_1^0}
			\right)
			+ \frac{\pi}{2}, \\
			\theta_\Gamma &=
			\arg \Gamma(1+\nu), \qquad
			\theta_\Gamma^{\rm as}
			= |\nu| - |\nu|\ln|\nu| - \frac{\pi}{4}.
		\end{aligned}
	\end{equation}
	where $\Gamma(1+\nu)$ is Euler’s gamma function~\cite{FialkovskyPerel}.
	
	We note that formula (52) from \cite{FialkovskyPerel} contains a misprint, however (93) and comments between (92) and (93) are correct.

	\subsection{Transfer Matrices ${\mathcal T}^0$ and ${\mathcal T}$}
	
	To apply the transfer matrix \eqref{eq:transfer-real-v-0}, we first determine the eigenvalues and eigenfunctions of the unperturbed problem \eqref{eq:unpert-0}. A straightforward calculation gives
	\begin{eqnarray}\label{eq:num-np}
		& \beta_{j}^0(x) =& (-1)^j \, \sign(x-\varkappa_{\E}) \sqrt{(U(x)-U(\varkappa_{\E}))^2 - (A_y(x)-A_y(\varkappa_{\E}))^2},\\
		& \boldsymbol{\vph}_{j}^0(x) =& \begin{pmatrix} \beta_{j}^0(x) - i (A_y(x) - A_y(\varkappa_{\E}))\\ \E - U(x) \end{pmatrix} \frac{1}{x-\varkappa_{\E}}. \label{eq:phi-pm}
	\end{eqnarray}
	Here, $\boldsymbol{\vph}_{j}^0(x)$ is defined up to a scalar function of $x$, which we choose as $\frac{1}{x-\varkappa_{\E}}$ to ensure a finite  limit as $x \to \varkappa_{\E}$. The eigenvalue indices are assigned consistently with their behavior near $x = \varkappa_{\E}$, see \eqref{eq:eigenvalues-difference-0}.
	
	Near $x \to \varkappa_{\E}$, using
	\begin{equation}
		\E - U(x) \approx -U'(\varkappa_{\E}) (x - \varkappa_{\E}), \quad A_y(x) - A_y(\varkappa_{\E}) \approx B(\varkappa_{\E}) (x - \varkappa_{\E}),
	\end{equation}
	we obtain
	\begin{eqnarray}
		& \beta_{j}^0(x) =& (-1)^j \, Q(\varkappa_{\E}) (x - \varkappa_{\E}) + o(x - \varkappa_\E), \label{eq:eta}\\
		& Q(\varkappa_\E) =& \sqrt{(U'(\varkappa_{\E}))^2 - B^2(\varkappa_{\E})}, \label{eq:def-Q}\\
		& \boldsymbol{\vph}_{j}^0(\varkappa_{\E}) =& \begin{pmatrix} (-1)^j Q(\varkappa_\E) - i B(\varkappa_{\E})\\ - U'(\varkappa_{\E}) \end{pmatrix} + \ldots \label{eq:phi-to-dp}
	\end{eqnarray}
	The vectors $\boldsymbol{\vph}_{j}^0(\varkappa_{\E})$, $j=1,2$, are linearly independent and smooth in $x$, fulfilling all conditions required for the application of the transfer matrix formalism from \cite{FialkovskyPerel}.
	
	For comparison of ${\boldsymbol{\Psi}}_j^{c}$ with the simplified semiclassical solutions used in the definition of ${\mathcal T}^0$, we expand
	\begin{equation}\label{eq:for_comparison}
		{\cal P}(x) = A_y(x) - A_y(\varkappa_\E) + \sqrt{\hbar} {\cal P}_{\E} \approx B(\varkappa_{\E}) (x - \varkappa_{\E}) + \sqrt{\hbar} {\cal P}_{\E}.
	\end{equation}
	Outside the classically allowed region, $|x - \varkappa_{\E}| \ge \hbar^{1/2 - \alpha'}$ with $0 < \alpha' < 1/6$ \cite{FialkovskyPerel}, the term $\sqrt{\hbar}\,{\cal P}_{\E}$ can be neglected at leading order in $\hbar$, giving
	\begin{eqnarray}\label{eq:eigen-compare}
		\beta_j \approx \beta_j^0, \quad
		\frac{\boldsymbol{\varphi}_{j}(x)}{|N_j(x)|^{1/2}}
		\approx
		\operatorname{sign}(x-\varkappa_{\E})
		\frac{\boldsymbol{\varphi}_{j}^0(x)}{|N_j^0(x)|^{1/2}},
		\quad
		\gamma_j(x) \approx \gamma_j^0(x),
	\end{eqnarray}
	where, according to \eqref{eq:Berry},
	\begin{equation}
		\gamma_j^0(x) =
		\frac{(-1)^{j+1}}{2}
		\,\operatorname{sign}(x-\varkappa_{\E})
		\arcsin
		\frac{A_y(x)-A_y(\varkappa_{\E})}
		{|U(x)-U(\varkappa_{\E})|}
	\end{equation}
	and 	
	\begin{equation}
		-\int_{\varkappa_c}^x \Im S_j^0(x') \, dx' = -\gamma_j^0(x) + \gamma_j^0(\varkappa_c).
	\end{equation}
	At the lower integration limit,
	\begin{equation}\label{eq:P-simp_low}
		\left. \arcsin \frac{A_y(x) - A_y(\varkappa_{\E})}{|\E - U(x)|} \right|_{x = \varkappa_c} \approx \arcsin \frac{B(\varkappa_{\E})}{|U'(\varkappa_{\E})|} \, \sign(\varkappa_c - \varkappa_{\E}).
	\end{equation}
	
	Finally, the semiclassical solutions can be written as
	\begin{multline}\label{eq:Psi-j}
		\boldsymbol{\Psi}_j^{c,0}(x) = \frac{\boldsymbol{\vph}_{j}^0(x)}{|N_j^0(x)|^{1/2}} \\ 
		\cdot \exp \Bigg\{ i (-1)^j \sign(x - \varkappa_{\E}) \Big[ \int_{\varkappa_c}^x \frac{|\beta_j^0(x')|}{\hbar} dx' + \frac{1}{2} \arcsin \frac{A_y(x) - A_y(\varkappa_{\E})}{|U(x) - U(\varkappa_{\E})|} \Big] + i \gamma_{j}^0(\varkappa_c)\, \Bigg\},
	\end{multline}
	where
	\begin{equation}\label{eq:chi}
		\gamma_{j}^0(\varkappa_c)  = (-1)^{j+1}\chi^0_{\E},\quad\chi^0_{\E} = \frac{1}{2} \arcsin \frac{B(\varkappa_{\E})}{|U'(\varkappa_{\E})|}.
	\end{equation}
	
	The leading-order relation between solutions with and without the zero superscript is
	\begin{equation}\label{eq:hat-no-hat}
		\boldsymbol{\Psi}_j^{c}(x) \approx \sign(x - \varkappa_{\E}) \boldsymbol{\Psi}_j^{c,0}(x) \exp \Big[ i (-1)^j (\chi^0_{\E} - \sign(x - \varkappa_{\E}) \chi_{\E}) \Big],\qquad c=\pm,
	\end{equation}
	see \eqref{eq:chi}, \eqref{eq:gamma-E}, \eqref{eq:betas}. 
	Consequently, the semiclassical solutions differ solely by a sign—stemming from the amplitude in  \eqref{eq:eigen-compare}—and by a constant phase factor, which reflects the choice of the lower integration limit in the Berry phase.

	The transfer matrix ${\cal T}^0$, defined in \eqref{eq:transfer-real-v-0}, is expressed in terms of the parameters $\nu$ and $\Theta_1$. Using \eqref{eq:phi-to-dp} and \eqref{for_matrix-0}, we calculate
	\begin{align}
		& {\mathcal B}_{12}^{(0)} = -2 i {\cal P}_{\E} U'(\varkappa_{\E}) (-Q(\varkappa_{\E}) + i B(\varkappa_{\E})), \quad
		N_1^{(0)} = 2 Q(\varkappa_{\E}) U'(\varkappa_{\E}), \quad N_2^{(0)} = - N_1^{(0)}, \label{eq:A-choice2-B12} \\
		& \frac{{\mathcal B}_{12}^{(0)}}{N_1^{(0)}} = i {\cal P}_{\E} \left(1 - \frac{i B(\varkappa_{\E})}{Q(\varkappa_{\E})}\right), \quad
		{\mathcal B}_{21}^{(0)} = \overline{{\mathcal B}_{12}^{(0)}}. \label{eq:B12-N1}
	\end{align}
	
	The controlling parameter $\nu$ is
	\begin{equation}\label{eq:sigma2}
		\nu = -i \frac{(p_y + A_y(\varkappa_{\E}))^2}{2 \hbar |U'(\varkappa_{\E})|} \frac{1}{\left(1 - (B(\varkappa_{\E})/|U'(\varkappa_{\E})|)^2\right)^{3/2}}.
	\end{equation}
	
	The phase factor $\Theta_1$ reads
	\begin{equation}\label{eq:Theta1-mu-chi}
		\Theta_1 = (\chi_{\E} - \chi^0_{\E}) + \frac{1}{2} \left( \frac{\pi}{2} + \vartheta_{\Gamma} - \vartheta_{\Gamma}^{as} \right),
	\end{equation}
	with $\chi_{\E}$ and $\chi^0_{\E}$ defined in   \eqref{eq:gamma-E}, \eqref{eq:chi}.
	
	Finally, the transfer matrix ${\cal T}$ is written as
	\begin{equation}\label{eq:transfer-hat-no-hat}
		{\cal T} = - \begin{pmatrix}
			e^{ -2 i \chi_{\E}}	{\cal T}^0_{11} & e^{ 2 i \chi_{\E}^0}{\cal T}^0_{12} \\
			e^{ -2 i \chi_{\E}^0}	{\cal T}^0_{21} & e^{ 2 i \chi_{\E}}{\cal T}^0_{22}
		\end{pmatrix}.
	\end{equation}

	\section{ Reflection and Transmission}
	
	\subsection{Single $n$-$p$ and $p$-$n$ junctions}
	
	The reflection and transmission coefficients are expressed in terms of the entries of the transfer matrix
	$\mathcal T_{\text{n-p}} = \mathcal T$ (see  \eqref{eq:r-t-matrix} and \eqref{eq:T-defin}).  Using \eqref{eq:transfer-hat-no-hat}, we obtain
	\begin{equation}\label{eq:hat-refl-trans}
		\begin{aligned}
			&r_{\text{n-p}} = e^{-2i(\chi_{\E}^0 + \chi_{\E})}\, r_{\text{n-p}}^{0}, \quad
			& r_{\text{n-p}}^{0} &=
			- \frac{{\mathcal T}^0_{21}}{\mathcal T^0_{22}}, \\
			&t_{\text{n-p}} = -e^{-2i\chi_{\E}}\, t_{\text{n-p}}^{0}, \quad
			&t_{\text{n-p}}^{0} &=
			\frac{1}{\mathcal T^0_{22}},
		\end{aligned}
	\end{equation}
	where ${\mathcal T}_{\text{n-p}}^0 = {\mathcal T}^0$ and $\det \mathcal T_{\text{n-p}}^0 = 1$.
	
	Equations \eqref{eq:hat-refl-trans} and \eqref{eq:transfer-real-v-0} yield
	\begin{equation}\label{eq:res-refl-trans-A}
		\begin{aligned}
			r_{\text{n-p}}^{0} = - e^{-2i \Theta_1} \sqrt{1 - e^{-2 \pi |\nu|}}, \quad
			t_{\text{n-p}}^{0} = e^{- \pi |\nu|}.
		\end{aligned}
	\end{equation}
	Taking into account \eqref{eq:Theta1-mu-chi}, we obtain
	\begin{equation}\label{eq:res-refl-trans-A2}
		\begin{aligned}
			r_{\text{n-p}}^{0} = - e^{-2i (\chi_{\E} -\chi_{\E}^0) - i \left(\frac{\pi}{2}
				+  \vartheta_{\Gamma} - \vartheta_{\Gamma}^{as}\right)} 
			\sqrt{1 - e^{-2 \pi |\nu|}}, \quad
			t_{\text{n-p}}^{0} = e^{- \pi |\nu|}.
		\end{aligned}
	\end{equation}
	
	Using the relation $e^{-2i\chi_{\E} - i \pi/2} = \sign({\cal P}(\varkappa_\E))$ for ${\cal P}(\varkappa_\E)\neq 0$, we obtain
	\begin{equation}\label{eq:res-refl-trans-fin}
		r_{\text{n-p}}^{0} =
		\sign\!\left(p_y + A_y(\varkappa_{\E})\right)\,
		e^{ i\arctan\!\left[\frac{B(\varkappa_{\E})}{Q(\varkappa_{\E})}\right]
			- i (\vartheta_{\Gamma} - \vartheta_{\Gamma}^{as}) }
		\sqrt{1 - e^{-2 \pi |\nu|}} .
	\end{equation}
	The reflection coefficient changes sign upon reversal of sign of ${\cal P}(\varkappa_\E)$, in agreement with the prediction of \cite{shytov2008klein} for $B=0$.  The solutions $\boldsymbol{\Psi}_j^{c,0}(x)$ remain smooth as functions of ${\cal P}(\varkappa_\E)$.
	If ${\cal P}(\varkappa_\E)=0$, then $\nu=0$ and $r_{\text{n-p}}^{0}=0$ according to  \eqref{eq:res-refl-trans-fin}; hence its phase is undefined.

	For $|\nu|\gg 1$, we have $\vartheta_{\Gamma} - \vartheta_{\Gamma}^{as} \ll 1$, and therefore
	\begin{equation}\label{eq:res-refl-trans-large-nu}
		r_{\text{n-p}}^{0} \approx
		\sign\!\left(p_y + A_y(\varkappa_{\E})\right)\,
		e^{ i\arctan\!\left[\frac{B(\varkappa_{\E})}{Q(\varkappa_{\E})}\right] }.
	\end{equation}
	
	In the opposite limit $|\nu|\ll 1$, we have $\vartheta_{\Gamma} - \vartheta_{\Gamma}^{as} \approx \pi/4$, and
	\begin{equation}
		r_{\text{n-p}}^{0} \approx
		\sqrt{\frac{\pi}{ |U'(\varkappa_{\E})|}}
		\frac{{\cal P}_{\E}}{(1-(B(\varkappa_{\E})/U'(\varkappa_{\E}))^2)^{3/4}}
		e^{ i\arctan\!\left[\frac{B(\varkappa_{\E})}{Q(\varkappa_{\E})}\right] - i\pi/4 },
		\quad
		{\cal P}_{\E} \equiv \frac{{\cal P}(\varkappa_\E)}{\sqrt{\hbar}},
	\end{equation}
	where we used \eqref{eq:sigma2} for $\nu$ and the definition of ${\cal P}_{\E}$,
	see equations \eqref{eq:gamma-def-0} and \eqref{eq:del-h-0}.
	By substituting \eqref{eq:res-refl-trans-A2}  into  \eqref{eq:hat-refl-trans}, we obtain    
	\begin{equation}\label{eq:hat-refl-trans-np}
		r_{\text{n-p}} =
		e^{- i \left(\frac{\pi}{2}
			+  \vartheta_{\Gamma} - \vartheta_{\Gamma}^{as}\right)}
		\sqrt{1 - e^{-2 \pi |\nu|}},
		\qquad
		t_{\text{n-p}} = -e^{-i\frac{\pi}{2} \sign{\cal P}_{\E}} e^{- \pi |\nu|},
	\end{equation}
	where we used $e^{-4i\chi_{\mathcal{E}}} = -1$, as it follows from \eqref{eq:gamma-E}. For $\nu=0$, $t_{\text{n-p}}=-1$, since the right-moving semiclassical solutions acquire a sign change when crossing $\varkappa_{\mathcal{E}}$ (equations \eqref{eq:hat-no-hat}, \eqref{eq:solutions-corresp}).


	Reflection and transmission coefficients from the $p$-$n$ junction are determined in \eqref{eq:r-t-defin-pn}. We take into account that ${\mathcal T}_{p-n} = \tilde{\mathcal T},$ $\det{\tilde{\mathcal T}} =1.$ The matrix $\tilde{\mathcal T}^0$ is determined by the formula \eqref{eq:transfer-real-v-0}
	where in the definitions of $\nu$ and $\Theta_1$ given in \eqref{eq:sigma2} and \eqref{eq:Theta1-mu-chi},
	$\varkappa_{\E}$ is replaced by $\widetilde\varkappa_{\E}.$ Equation \eqref{eq:transfer-hat-no-hat}, where ${\chi}_{\E}^0$ and $\chi_{\E}$ are replaced  with $\widetilde{\chi}_{\E}^0$ and $\widetilde{\chi}_{\E}$, expresses $\widetilde{\mathcal T}$ through $\widetilde{\mathcal T}^0$ .   We 
	get
	\begin{equation}
		\begin{aligned}
			&r_{\text{p-n}} = e^{2i(\tilde{\chi}_{\E}^0 + \tilde\chi_{\E}) }r_{\text{p-n}}^0, \quad 
			& r_{\text{p-n}}^0  &= - \frac{\widetilde{\mathcal T}_{12}^0}{\widetilde{\mathcal T}_{11}^0}, \\ &t_{\text{p-n}} = -e^{2i\tilde\chi }t_{\text{p-n}}^0,  \quad 
			&t_{\text{p-n}}^0 &= \frac{1}{\widetilde{\mathcal T}_{11}^0},
		\end{aligned}
	\end{equation}
	Denoting $\nu$, $\vartheta_{\Gamma}$, $\vartheta_{\Gamma}^{as}$, calculated using $\widetilde\varkappa_{\E}$, with the tilde symbol, we obtain
	\begin{equation}\label{eq:hat-refl-trans-pn}
		\begin{aligned}
			&	r_{\text{p-n}}^{0} = - e^{2i (\tilde\chi -\tilde\chi_{\E}^0) + i \left(\frac{\pi}{2}
				+  \tilde\vartheta_{\Gamma} - \tilde\vartheta_{\Gamma}^{as}\right)} 
			\sqrt{1 - e^{-2 \pi |\tilde\nu|}}, \qquad
			&t_{\text{p-n}}^{0} = e^{- \pi |\tilde\nu|},\qquad\qquad\\
			& r_{\text{p-n}} = e^{ i (\frac{\pi}{2}
				+  \tilde{\theta}_{\Gamma} - \tilde{\theta}_{\Gamma}^{as}) }\sqrt{1 - e^{-2 \pi \vert\tilde{\nu}\vert}}, \quad  &t_{\text{p-n}} = -e^{i\frac{\pi}{2} \sign{\widetilde{\cal P}_{\E}}}e^{- \pi \vert\tilde{\nu}\vert}.
		\end{aligned}
	\end{equation}
	
	The transmission coefficients for a single $n\text{–}p$ or $p\text{–}n$ interface are controlled by the parameter $\nu$ (or $\tilde{\nu}$). Using \eqref{eq:py}, perfect transmission occurs for incidence at the angle $\theta_0(\E)$ defined by
	\begin{equation}
		\sin\theta_0 \equiv -\frac{A_y(\varkappa_{\E})}{\E},
	\end{equation}
	assuming $\theta_0$ is real. In this case
	\begin{equation}\label{eq:nu-dim}
		\nu = -i \frac{\E^2}{2\hbar |U'(\varkappa_{\E})|}
		(\sin\theta-\sin\theta_0)^2
		\left(1-\left(\frac{B(\varkappa_{\E})}{U'(\varkappa_{\E})}\right)^2\right)^{-3/2}.
	\end{equation}
	In the semiclassical regime $\E^2/(2\hbar |U'|)\gg1$, transmission is confined to $|\theta-\theta_0|\ll1$. Expanding \eqref{eq:nu-dim} to leading order in $\theta-\theta_0$, we obtain
	\begin{equation}
		\nu \approx
		-i \frac{\E^2}{2\hbar |U'(\varkappa_{\E})|}
		(\theta-\theta_0)^2
		\frac{
			1-
			\left(
			\displaystyle
			\frac{\int_{x_*}^{\varkappa_{\E}} B(\tilde x)\, d\tilde x}
			{\int_{x_*}^{\varkappa_{\E}} U'(\tilde x)\, d\tilde x}
			\right)^2
		}
		{
			\left(1-\left(\frac{B(\varkappa_{\E})}{U'(\varkappa_{\E})}\right)^2\right)^{3/2}
		}.
	\end{equation}

	\subsection{Fabry--P\'erot resonances}
	
	To evaluate the transmission coefficient through an $n$-$p$-$n$ junction using
	\eqref{eq:trans-npn}, we express the transfer matrix of the $n$-$p$-$n$
	structure (see \eqref{eq:product-transfer}) in terms of the reflection and
	transmission coefficients of the $n$-$p$ and $p$-$n$ junctions.
	The corresponding transfer matrices
	${\mathcal T}_{\text{n-p}}$ and
	${\mathcal T}_{\text{p-n}}$
	can be written as
	\begin{equation}\label{eq:matr-n-p-coef}
		{\mathcal T}_{\text{n-p}} =
		\left(\begin{array}{cc}
			1/\overline{t_{\text{n-p}}} &
			-\overline{r_{\text{n-p}}}/\overline{t_{\text{n-p}}} \\
			- r_{\text{n-p}}/t_{\text{n-p}} &
			1/t_{\text{n-p}}
		\end{array}\right),\qquad
		{\mathcal T}_{\text{p-n}} =
		\left(\begin{array}{cc}
			1/t_{\text{p-n}} &
			- r_{\text{p-n}}/t_{\text{p-n}} \\
			- \overline{r_{\text{p-n}}}/\overline{t_{\text{p-n}}} &
			1/\overline{t_{\text{p-n}}}
		\end{array}\right).
	\end{equation}
	These expressions follow from \eqref{eq:r-t-matrix} and
	\eqref{eq:r-t-defin-pn}.
	After straightforward calculations, \eqref{eq:trans-npn} yields
	\begin{equation}\label{eq:Tnpn}
		t_{\text{n-p-n}} =
		\frac{
			|t_{\text{n-p}}|\, |t_{\text{p-n}}|
			e^{- i (\zeta - \arg t_{\text{n-p}} - \arg t_{\text{p-n}}  )}
		}
		{
			1 + e^{-2 i (\zeta - \arg t_{\text{n-p}})}
			\overline{r_{\text{n-p}}}\, r_{\text{p-n}}
		}.
	\end{equation}
	
	The Fabry--P\'erot resonances are determined by the condition
	\begin{equation}\label{eq:denom}
		1 + e^{- 2 i (\zeta - \arg t_{\text{n-p}})}
		\overline{r_{\text{n-p}}}\, r_{\text{p-n}} = 0 .
	\end{equation}
	Substituting into \eqref{eq:denom} the expressions for $\zeta$,
	\eqref{eq:zeta}, $t_{\text{n-p}}$ and $r_{\text{n-p}},$
	see \eqref{eq:hat-refl-trans-np}, and $r_{\text{p-n}},$
	see \eqref{eq:hat-refl-trans-pn}, we obtain
	\begin{equation}\label{eq:Fabry-P}
		\delta(\E,\theta) = 2 \pi M + i \ln |r_{\text{n-p}} r_{\text{p-n}}| ,
	\end{equation}
	where $M$ is an integer and
	\begin{equation}\label{eq:delta}
		\delta(\E,\theta) \equiv
		-\frac{2}{\hbar}
		\int\limits^{\tilde{\varkappa}_{h}}_{{\varkappa}_{h}}
		|\beta(x')|\, dx'
		- \frac{\pi}{2}
		\bigl(
		\sign{\mathcal P_{\E}} +
		\sign{\tilde{\mathcal P}_{\E}}
		\bigr)
		+ \vartheta_{\Gamma} - \vartheta_{\Gamma}^{\mathrm{as}}
		+ \tilde{\theta}_{\Gamma} - \tilde{\theta}_{\Gamma}^{\mathrm{as}} .
	\end{equation}
	
	Following \cite{Tudorovskiy_2012,reijnders2013semiclassical}, we calculate
	$|t_{\text{n-p-n}}|^2$ as a function of the angle of incidence of the charge carrier.
	Multiplying \eqref{eq:Tnpn} by its complex conjugate, we find
	\begin{equation}\label{eq:Tnpn-modul}
		|t_{\text{n-p-n}}|^2 =
		\frac{
			|t_{\text{n-p}} t_{\text{p-n}}|^2
		}
		{
			(1 - |r_{\text{n-p}} r_{\text{p-n}}|)^2
			+ 4 |r_{\text{n-p}} r_{\text{p-n}}|
			\sin^2(\delta/2)
		}.
	\end{equation}

	\begin{figure}[h]
		\centering
		\begin{subfigure}[b]{0.33\textwidth}
			\centering
			\includegraphics[width=\linewidth]{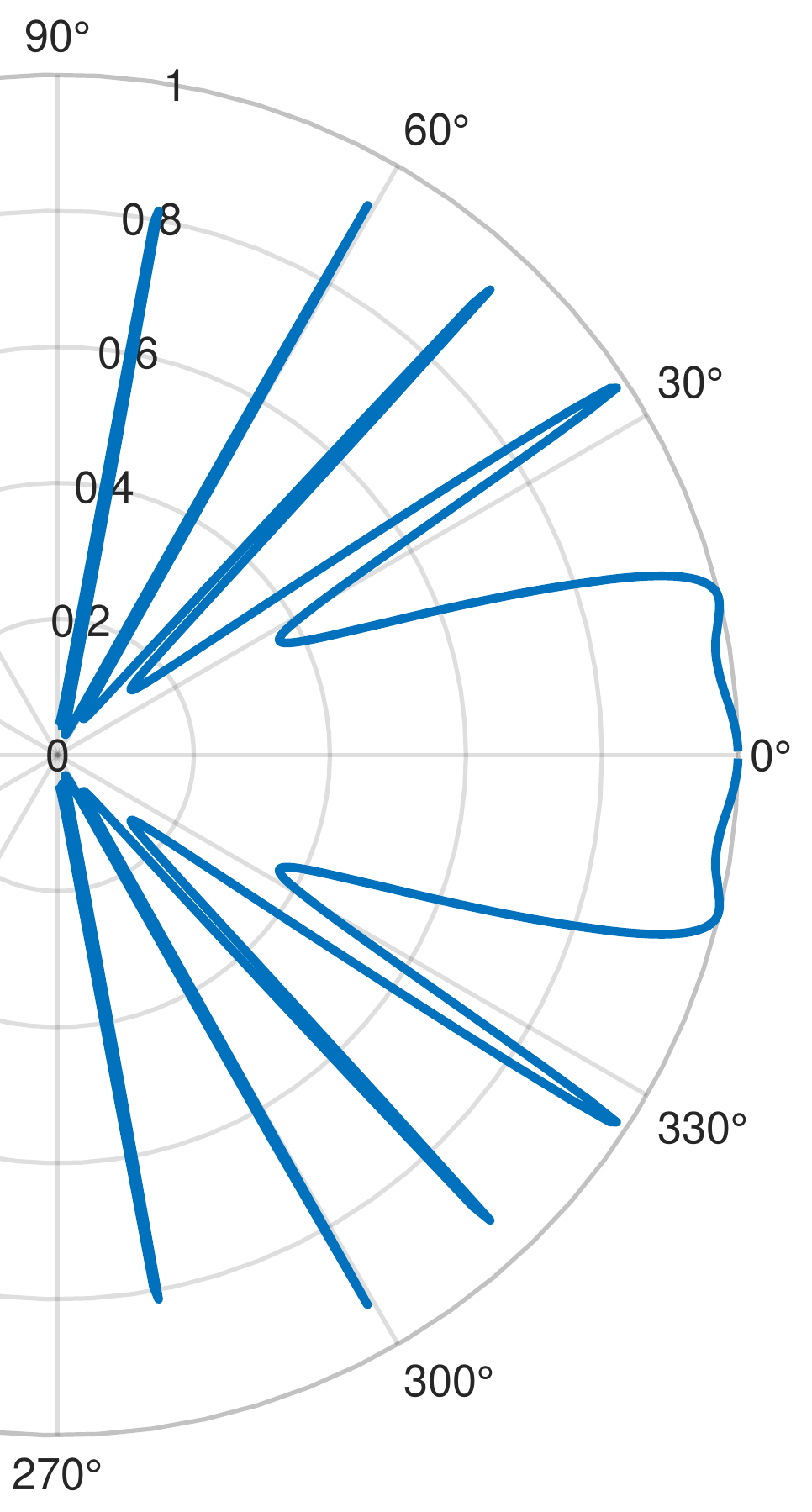}
			\caption{No magnetic field}
			\label{fig:n0}
		\end{subfigure}%
		\begin{subfigure}[b]{0.33\textwidth}
			\centering
			\includegraphics[width=\linewidth]{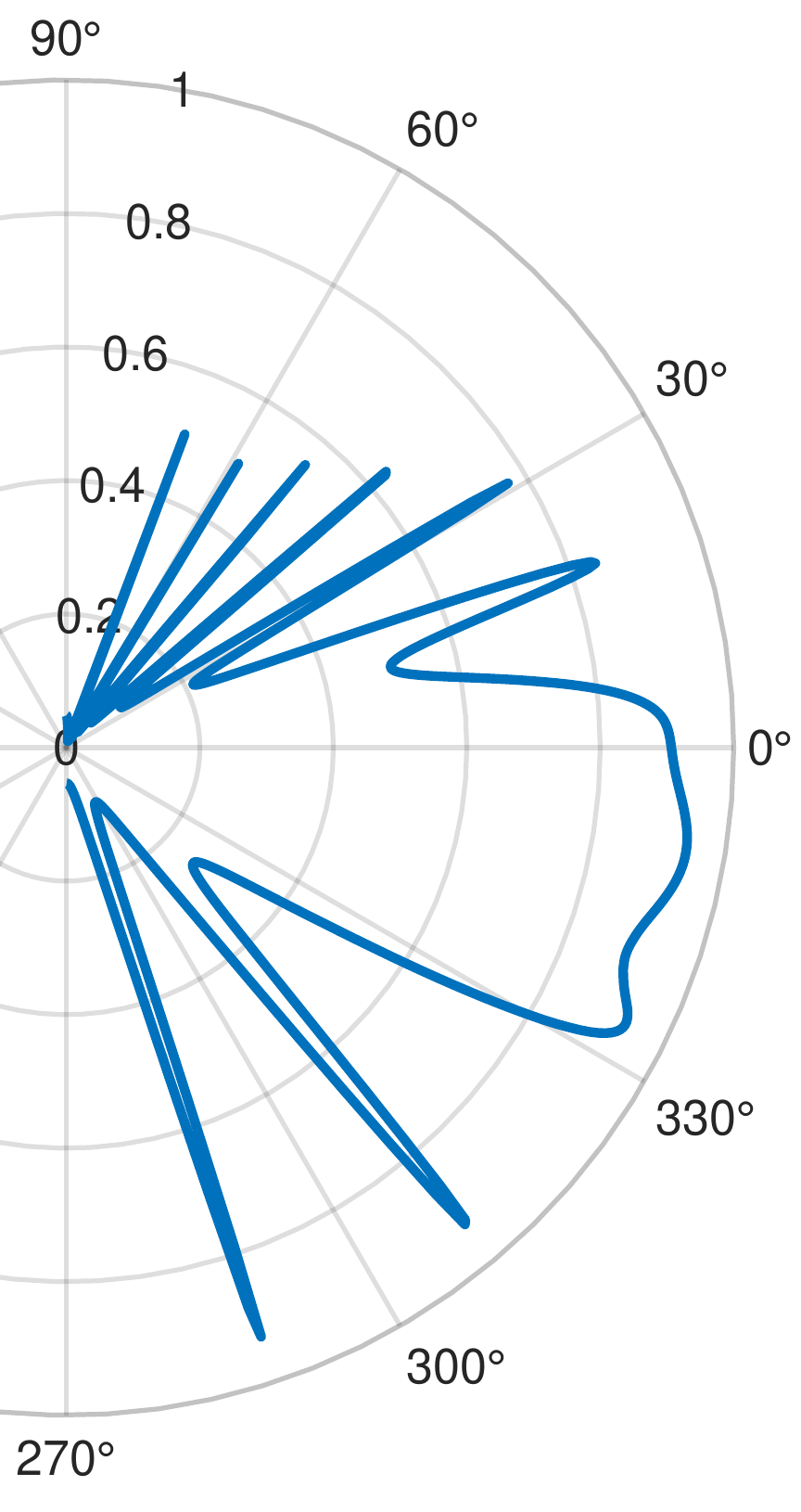}
			\caption{Magnetic field shown by the dotted line in Fig.~\ref{Fig1}(c)}
			\label{fig:n2}
		\end{subfigure}%
		\begin{subfigure}[b]{0.33\textwidth}
			\centering
			\includegraphics[width=\linewidth]{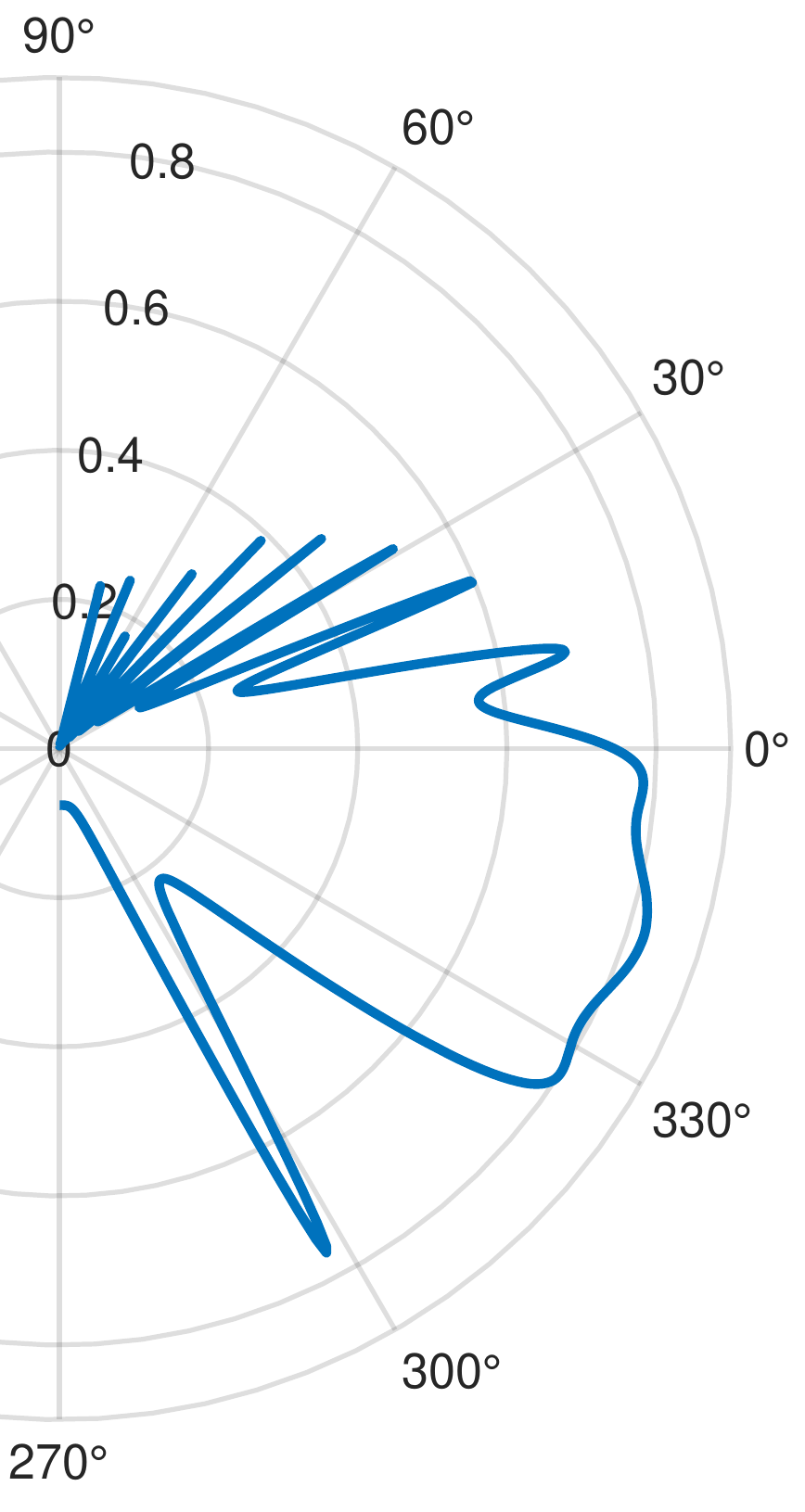}
			\caption{Magnetic field shown by the solid line in Fig.~\ref{Fig1}(c)}
			\label{fig:n1}
		\end{subfigure}
		\caption{The absolute value of $t_{\text{n-p-n}}$ as a function of the angle of incidence on the barrier.}
		\label{fig:Tnpn-versus-theta}
	\end{figure}
	
	Figure~\ref{fig:Tnpn-versus-theta} shows the absolute value of
	$t_{\text{n-p-n}}$ as a function of the angle of incidence $\theta$.
	Equation~\eqref{eq:Tnpn-modul} demonstrates that the transmission probability
	reaches its maximum when $\delta = 2 \pi M$, where $M$ is an integer.
	The corresponding angles are referred to as ``magic'' angles
	\cite{katsnelson2006chiral,Tudorovskiy_2012,reijnders2013semiclassical}.
	The maximal transmission satisfies $|t_{\text{n-p-n}}|_{\max} = 1$ if
	$|r_{\text{n-p}}| = |r_{\text{p-n}}|$, therefore
	suppression of transmission
	requires an asymmetric barrier.
	Using \eqref{eq:Tnpn-modul}, we have calculated the transmission probability
	through the barrier shown in Fig.~\ref{Fig1}(b) with magnetic-field profiles
	depicted in Fig.~\ref{Fig1}(c).
	Figure~\ref{fig:Tnpn-versus-theta} presents a polar plot of
	$|t_{\text{n-p-n}}|$ versus the angle of incidence $\theta$.
	In the absence of a magnetic field
	[Fig.~\ref{fig:Tnpn-versus-theta}(a)],
	the positions of the “magic” angles coincide with those reported in
	\cite{reijnders2013semiclassical}.
	An external magnetic field introduces additional asymmetry, resulting in a shift
	of the “magic” angle positions.
	If 
	$U(-x)=U(x)$ and $B(-x)=-B(x)$, full transmission at the “magic” angles remains possible even in the presence of a magnetic field; in this case 
	$A_y(-x)=A_y(x)$ and $\nu(\varkappa_{\E})=\nu(\tilde{\varkappa}_{\E})$.
	
	We anticipate a qualitative modification of the interference structure along the lines in the $(p_y,\E)$ plane defined by
	\begin{equation}
		{\cal P}(\varkappa_{\E})=0, \qquad {\cal P}(\tilde\varkappa_{\E})=0,
	\end{equation}
	which correspond to
	\begin{equation}\label{eq:p1-2-A}
		p_y^{(1)} = -A_y(\varkappa_{\E}), \qquad
		p_y^{(2)} = -A_y(\tilde\varkappa_{\E}),
	\end{equation}
	where $|t_{\text{n-p}}|=1$ and $|t_{\text{p-n}}|=1$, respectively.
	At these lines one of the junctions becomes perfectly transparent, so that the corresponding reflection coefficient vanishes and changes sign upon crossing. As follows from \eqref{eq:delta}, this sign reversal produces a discontinuous jump of the phase $\delta$ by $\pi$.		
	Since $\delta=2\pi M$ corresponds to constructive interference and $\delta=(2M+1)\pi$ to destructive interference, the $\pi$ phase jump interchanges transmission maxima and minima. Thus, crossing either of these lines in the $(p_y,\E)$ plane results in an abrupt reconfiguration of the interference pattern.

	\section{Conclusion}
	
	In this work, we developed a semiclassical description of fermion tunneling through an inhomogeneous potential barrier in graphene in the presence of an inhomogeneous magnetic field. Both the electrostatic and magnetic fields are assumed to vary smoothly along the transport ($x$) direction while remaining uniform in the transverse ($y$) direction. Within this framework, we derived the magnitudes and phases of the reflection and transmission coefficients for $n$–$p$ and $p$–$n$ junctions; the corresponding results are given in \eqref{eq:hat-refl-trans-np} and \eqref{eq:hat-refl-trans-pn}. In the absence of a magnetic field, our results are consistent with those of \cite{Tudorovskiy_2012,reijnders2013semiclassical}.
	
	We further analyzed electron transmission through an $n$–$p$–$n$ junction. Analytical expressions for the Fabry–Pérot resonances and the so-called “magic” angles were obtained. We predict a restructuring of the interference pattern at angles of incidence and energies corresponding to a complete transition through one of the junctions $n$-$p$ or $p$-$n$.
	
	In addition, we performed numerical calculations of the transmission coefficient as a function of the angle of incidence. In the limiting case of a vanishing magnetic field, our results reduce to those reported in \cite{reijnders2013semiclassical}, demonstrating full agreement.
	\vskip6pt
	
	\enlargethispage{20pt}

	\appendix
	\section{Zero–Magnetic-Field Limit}
	
	In this Appendix, we compare the semiclassical expressions obtained in the present work for the reflection and transmission coefficients $r_{\text{n-p}}$ and $t_{\text{n-p}}$ with the results of \cite{Tudorovskiy_2012,reijnders2013semiclassical} in the absence of a magnetic field. Specifically, we relate the semiclassical modes given in \eqref{eq:solutions-corresp},\eqref{eq:adiab-last} to the asymptotic scattering states (C.3) and (C.6) of \cite{reijnders2013semiclassical}.
	
	We adopt the notation $\phi_p^+(x)$ and $v(x)$ introduced in (C.4) of  \cite{reijnders2013semiclassical}:
	\begin{equation}
		\cos{\phi_p^+(x)}= \frac{\beta(x)}{|v(x)|}, \quad 
		\sin{\phi_p^+(x)}= \frac{{\cal P}(x)}{|v(x)|}, \quad 
		U(x)-{\cal E}=v(x).
	\end{equation}
	Here $\beta(x)$ is defined in \eqref{eq:beta} and satisfies $\beta(x)>0$. It follows that
	\begin{equation}
		\beta^2(x)+{\cal P}^2(x)=v^2(x), \qquad 
		\sqrt{\beta^2(x)+{\cal P}^2(x)}=|v(x)|.
	\end{equation}
	Consequently,
	\begin{equation}
		\begin{aligned}
			& \phi_p^+(x)=\arcsin{\frac{{\cal P}(x)}{|{\cal E}-U(x)|}}, \\
			& \beta(x)\mp i{\cal P}(x)=|v(x)|e^{\mp i\phi_p^+(x)} .
		\end{aligned}
	\end{equation}
	
	In the electron region, where $|v(x)|=-v(x)$, the semiclassical solutions \eqref{eq:adiab-last}, \eqref{eq:solutions-corresp} take the form
	\begin{eqnarray}\label{eq:adiab-last-phi-e}
		\boldsymbol{\Psi}_{eR}(x) &=&
		\frac{|v(x)|}{|2\beta v(x)|^{1/2}}
		\begin{pmatrix}
			e^{-\frac{i}{2}\phi_p^+(x)}\\
			e^{\frac{i}{2}\phi_p^+(x)}
		\end{pmatrix}
		e^{ i\int^x_{\varkappa_e}\frac{\beta(x')}{\hbar}\,dx'
			- i\frac{\pi}{4}\sign{{\cal P}(\varkappa_{\cal E})}}, \\
		\boldsymbol{\Psi}_{eL}(x) &=&
		\frac{|v(x)|}{|2\beta v(x)|^{1/2}}
		\begin{pmatrix}
			- e^{\frac{i}{2}\phi_p^+(x)}\\
			e^{-\frac{i}{2}\phi_p^+(x)}
		\end{pmatrix}
		e^{- i\int^x_{\varkappa_e}\frac{\beta(x')}{\hbar}\,dx'
			+ i\frac{\pi}{4}\sign{{\cal P}(\varkappa_{\cal E})}} .
	\end{eqnarray}
	
	In the hole region, where $|v(x)|=v(x)$, one obtains
	\begin{eqnarray}\label{eq:adiab-last-phi-h}
		\boldsymbol{\Psi}_{hR}(x) &=&
		-\frac{|v(x)|}{|2\beta v(x)|^{1/2}}
		\begin{pmatrix}
			e^{\frac{i}{2}\phi_p^+(x)}\\
			e^{-\frac{i}{2}\phi_p^+(x)}
		\end{pmatrix}
		e^{- i\int^x_{\varkappa_e}\frac{\beta(x')}{\hbar}\,dx'
			+ i\frac{\pi}{4}\sign{{\cal P}(\varkappa_{\cal E})}}, \label{eq:Psi-hR} \\
		\boldsymbol{\Psi}_{hL}(x) &=&
		\frac{|v(x)|}{|2\beta v(x)|^{1/2}}
		\begin{pmatrix}
			e^{-\frac{i}{2}\phi_p^+(x)}\\
			- e^{\frac{i}{2}\phi_p^+(x)}
		\end{pmatrix}
		e^{ i\int^x_{\varkappa_e}\frac{\beta(x')}{\hbar}\,dx'
			- i\frac{\pi}{4}\sign{{\cal P}(\varkappa_{\cal E})}}. \label{eq:Psi-hL}
	\end{eqnarray}
	
	In the absence of a magnetic field, ${\cal P}(x)=p_y$ and $\sign{{\cal P}(\varkappa_{\cal E})}=\sign{p_y}$. For small $p_y$ changing sign, the semiclassical modes  $\boldsymbol{\Psi}_{ec},$ $\boldsymbol{\Psi}_{hc},$ $c=R, L$ exhibit a sharp variation due to the phase terms proportional to $\sign{p_y}$.

	We denote the semiclassical modes of \cite{reijnders2013semiclassical} by $\boldsymbol{\Psi}^{\mathrm{RTK}}_{\pm}(x)$. In that work, the phase $\phi_p^-(x)$ is defined such that
	\begin{equation}
		\phi_p^+(x)=\pi\sign{p_y}-\phi_p^-(x),
	\end{equation}
	which implies
	\begin{equation}
		\begin{aligned}
			- e^{\frac{i}{2}\phi_p^+(x)} &= - e^{ i\frac{\pi}{2}\sign{p_y}} e^{-\frac{i}{2}\phi_p^-(x)}, \\
			e^{-\frac{i}{2}\phi_p^+(x)} &= e^{- i\frac{\pi}{2}\sign{p_y}} e^{\frac{i}{2}\phi_p^-(x)} .
		\end{aligned}
	\end{equation}
	
	For $p_y\neq0$ and $x_0=\varkappa_e$, the electron-region solutions are related by
	\begin{equation}\label{eq:e-relation-TRK}
		\boldsymbol{\Psi}^{\mathrm{RTK}}_{+}(x)
		= e^{ i\frac{\pi}{4}\sign{p_y}} \frac{2}{\sqrt{|p_y|}} \boldsymbol{\Psi}_{eR}(x), \qquad
		\boldsymbol{\Psi}^{\mathrm{RTK}}_{-}(x)
		= e^{ i\frac{\pi}{4}\sign{p_y}} \frac{2}{\sqrt{|p_y|}} \boldsymbol{\Psi}_{eL}(x).
	\end{equation}
	
	The reflection coefficient $r_{\text{n-p}}$ defined in \eqref{eq:hat-refl-trans-np} coincides with the reflection coefficient $r$ of \cite{reijnders2013semiclassical}, given in (69) and (73) therein. The correspondence between parameters is
	\begin{equation}\label{eq:K-nu}
		\nu = - i \frac{K}{\pi \hbar}, \qquad
		\vartheta_{\Gamma}-\vartheta_{\Gamma}^{\mathrm{as}} = -\theta .
	\end{equation}
	
	The hole-region solutions (C.6) of   \cite{reijnders2013semiclassical} and solutions  \eqref{eq:Psi-hR}, \eqref{eq:Psi-hL} are related by
	\begin{equation}\label{eq:h-relation-TRK}
		\boldsymbol{\Psi}^{\mathrm{RTK}}_{+}(x)
		= e^{ i\frac{3\pi}{4}\sign{p_y}} \frac{2}{\sqrt{|p_y|}} \boldsymbol{\Psi}_{hL}(x), \qquad
		\boldsymbol{\Psi}^{\mathrm{RTK}}_{-}(x)
		= e^{ i\frac{3\pi}{4}\sign{p_y}} \frac{2}{\sqrt{|p_y|}} \boldsymbol{\Psi}_{hR}(x).
	\end{equation}
	This holds if equation (C.6) involves $S(x_0,x)$ rather than $S(x,x_0)$. Then, the transition coefficient $t$ from  \cite{reijnders2013semiclassical} (equation (62)) is related to $t_{\text{n-p}}$, as dictated by  \eqref{eq:h-relation-TRK} and \eqref{eq:e-relation-TRK}, i.e., $t=-e^{ i\frac{\pi}{2}\sign{p_y}} t_{\text{n-p}}$.

	\bibliographystyle{unsrt}
	\bibliography{tunnel_bib4}
	
\end{document}